# Nanothermodynamics: There's plenty of room on the inside

**Ralph V. Chamberlin [1,\*] and Stuart M. Lindsay [1,2]**

[1] Department of Physics, Arizona State University, Tempe, AZ USA 85287-1504; chamberl@asu.edu
[2] School of Molecular Science, Arizona State University, Tempe, AZ USA 85287-1604; stuart.lindsay@asu.edu
\* Correspondence: chamberl@asu.edu; Tel.: +1 (480) 965-3922 (RVC)

**Abstract:** Nanothermodynamics provides the theoretical foundation for understanding stable distributions of statistically independent subsystems inside larger systems. In this review it is emphasized that adapting ideas from nanothermodynamics to simplistic models improves agreement with the measured properties of many materials. Examples include non-classical critical scaling near ferromagnetic transitions, thermal and dynamic behavior near liquid-glass transitions, and the 1/$f$-like noise in metal films and qubits. A key feature in several models is to allow separate time steps for distinct conservation laws: one type of step conserves energy and the other conserves momentum (e.g. dipole alignment). This "orthogonal dynamics" explains how the relaxation of a single parameter can exhibit multiple responses such as primary, secondary, and microscopic peaks in the dielectric loss of supercooled liquids, and the crossover in thermal fluctuations from Johnson-Nyquist (white) noise at high frequencies to 1/$f$-like noise at low frequencies. Nanothermodynamics also provides new insight into three basic questions. First, it gives a novel solution to Gibbs' paradox for the entropy of the semi-classical ideal gas. Second, it yields the stable equilibrium of Ising's original model for finite-sized chains of interacting binary degrees of freedom ("spins"). Third, it confronts Loschmidt's paradox for the arrow of time, showing that an intrinsically irreversible step is required for maximum entropy and the second law of thermodynamics, not only in the thermodynamic limit but also in systems as small as $N = 2$ particles.

## 1. Introduction

Many measurements on most types of materials exhibit nanoscale heterogeneity. Although some of this heterogeneity is extrinsic – due to defects, impurities, etc. – here we focus on intrinsic heterogeneity that usually coexists with homogeneous behavior. This intrinsic heterogeneity comes from statistically independent subsystems that form inside most systems to minimize the free energy in the nanocanonical ensemble from nanothermodynamics, often yielding the largest and slowest fluctuations around equilibrium. Because the subsystems can be as small as a single atom, we have adapted the title of Feynman's seminal talk [1,2] to say there is plenty of room for subsystems inside most systems.

Standard thermodynamics starts by assuming that all systems are homogeneous and effectively infinite, so that an alternative approach is needed to treat the nanoscale fluctuations found in most materials. Similarly, standard statistical mechanics starts by assuming that local degrees of freedom have a well-defined temperature ($T$) from weak but essentially instantaneous coupling to a homogeneous and effectively infinite heat bath [3], whereas the concept of temperature is often ill-defined on the nanoscale [4-9]. In fact, various experimental techniques [10-14] and computer simulations [15,16] have shown that the effective local temperature ($T_i$) of an internal subsystem ($i$) can differ significantly from $T$, even during equilibrium fluctuations. This difference between $T_i$ and $T$ often arises when fluctuations in local energy occur faster than the time needed to couple them to the large heat bath. Fluctuation theorems and stochastic thermodynamics provide powerful tools for evaluating nanoscale fluctuations, especially under nonequilibrium conditions [17-19], but they usually require a canonical-ensemble distribution with at least one well-defined $T$, plus microreversibility during the dynamics that may violate the second law of thermodynamics, at least in simple models [16] and basic theory [20]. In any case, only the nanocanonical ensemble of nanothermodynamics gives the stability condition for thermal equilibrium of nanometer-sized subsystems. Here, we briefly review some of the theoretical justifications and experimental evidence for nanothermodynamics that provides the foundation for independent thermal fluctuations of nanoscale subsystems inside larger systems.

The remainder of this review is organized as follows. Section *2* starts with a short history of Hill's small-system thermodynamics [21-23]. Although Hill's ideas form the foundation of nanothermodynamics, we emphasize the conceptual differences and practical implications from systems that subdivide into independent internal subsystems, especially the crucial stability condition that is rarely (if ever) used in Hill's small-system thermodynamics. Section *3* gives an overview of six areas where nanothermodynamics has provided new insight. Section *3.1* gives a novel solution to Gibbs' paradox, resolving the discrepancy between classical statistical mechanics and thermodynamics for the measured entropies of dilute gases. Section *3.2* describes the stable equilibrium of the original Ising model for interacting binary degrees of freedom ("spins"). Such Ising-like models form the basis of most other applications in this review, including simplified models that provide improved agreement with measured non-classical critical scaling near ferromagnetic transitions (*3.3*), thermal and dynamic response near liquid-glass transitions (*3.4*), and 1/*f*-like noise in metal films, tunnel junctions, and qubits (*3.5*). Section *3.6* describes a simple model that utilizes the stable equilibrium of nanothermodynamics to investigate Loschmidt's paradox for the arrow of time. Thus, nanothermodynamics provides basic insight into multiple "unsolved problems in physics" [24]. Finally, section *4* gives a summary of the results and section *5* finishes with some conclusions.

**2. Background**

In 1962, Hill introduced the theory of small-system thermodynamics to facilitate the systematic treatment of finite-size thermal effects in large ensembles of small systems [21-23]. A novel result from this theory is a well-defined generalized ensemble that is completely open, having no extensive environmental variables, an ensemble that violates the usual Gibbs-Duhem relation of standard thermodynamics [25]. Here we emphasize how Hill's theory has been adapted to treat finite-size thermal effects from stable ensembles of small subsystems inside larger systems [25,26]. It is in the context of this adaptation that the term nanothermodynamics first appeared [27]. Similarly, the term nanocanonical ensemble was introduced for distributions of internal subsystems [28]. Later in this review we focus on experimental evidence for nanothermodynamics from many measurements, but first we discuss some general aspects of finite-size thermal effects.

The equation in the inset of Fig. 1 is Hill's fundamental equation (combined first and second laws) for small-system thermodynamics [22,25]. The first four terms (black) come from standard thermodynamics. The final term (red) contains Hill's subdivision potential ($\mathcal{E}$) and number of subdivisions ($\eta$), which are useful for small systems, and essential for the stable equilibrium of subsystems inside larger systems. (In general, the number of subdivisions is inversely proportional to subsystem size $\eta \sim N_t/n$.) Also shown is how this equation maps to a simple (three-energy-level) diagram for distinct contributions to conservation of energy during reversible processes. The first three-level diagram (left side of the equation) represents an initial distribution of energies. The next diagram (first on the right side) shows how adding heat changes the distribution of energies, without changing the levels. The next diagram (second on the right) shows how work done on the system changes the levels, without changing the distribution. The penultimate diagram shows how adding particles changes the total occupation of levels, without changing their distribution. The final diagram represents all non-extensive terms from finite-size effects, which may come from levels that are shifted due to higher-order corrections and/or broadened due to thermal fluctuations, surface states, interfaces, length-scale terms, etc. Standard thermodynamics, which requires simple systems to be effectively infinite and homogeneous, has no systematic way of including all these contributions needed to conserve total energy. Furthermore, the generalized ensemble (nanocanonical ensemble if applied to subsystems) is ill-defined without Hill's subdivision potential, analogous to how Gibbs' ensemble is ill-defined without Gibbs' chemical potential. In the generalized ensemble, where none of the extensive environmental variables is fixed, the fluctuations can be anomalously large. For example, individual small systems in contact with both a heat bath and a particle bath have mean-squared fluctuations in energy that increase with the square of the number of particles [29], not linearly as in standard thermodynamics, attributable to fluctuations in energy and in the number of particles. Such fluctuations can only be evaluated in the generalized ensemble of Hill's small-system thermodynamics. Similarly, fluctuations inside bulk systems require contributions from $\mathcal{E}$ if total energy is to be strictly conserved, especially on the scale of nanometers.

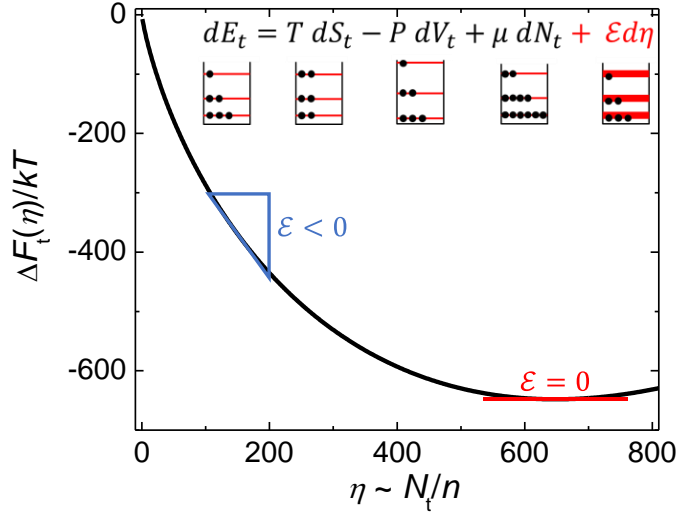

**Figure 1. Finite-size thermal effects.** Inset gives Hill's fundamental equation of small-system thermodynamics, with a simple (three-energy-level) diagram for each term. The first three terms on the right side (black) give the standard ways to increase the total internal energy of a system: add heat ($TdS_t > 0$), do work on the system ($-PV_t > 0$), or add particles ($\mu dN_t > 0$). The fourth term (red) contains finite-size effects (surface states, length-scale terms, fluctuations, etc.) that change the width of the levels when the number of subdivisions changes if the subdivision potential is nonzero ($\mathcal{E} \neq 0$). The main figure shows how free energy might change with the number of subdivisions, from $\Delta F_t(0) = 0$ in the thermodynamic limit of no subdivisions, to $\Delta F_t(0) = \min$ at $\mathcal{E} = 0$ in the nanothermodynamic limit for stable equilibrium of subsystems inside bulk samples.

The main part of Fig. 1 shows how the free energy of the 1D Ising model changes with the number of subdivisions, $\Delta F_t(\eta) = \int_0^\eta \mathcal{E} d\eta'$. The solid line shows $\Delta F_t(\eta)/kT = \eta \ln[\eta \cosh(J/kT)/N_t] - \eta$, found by integrating $\mathcal{E} = kT \ln[\eta' \cosh(J/kT)/N_t]$ (from Appendix B), using $J/kT = 1$ and $N_t = 10^3$ for specific values. Note that the equilibrium number of subdivisions is inversely proportional to the average subsystem size, $\bar{\eta} = N_t/\bar{n}$, so that $\bar{n} = \cosh(J/kT)$ gives thermal equilibrium, $\mathcal{E} = 0$. Large homogeneous systems ($\eta \to 0$) have $\Delta F_t(0) = 0$, while heterogeneous systems with internal subsystems have $\Delta F_t(\eta) < 0$. The slope, $\mathcal{E} = \partial \Delta F_t/\partial \eta|_{T,N_t}$, is negative as a large system starts to subdivide. Indeed, from Hill's fundamental equation in the inset of Fig. 1, total energy decreases whenever $\mathcal{E} d\eta < 0$. Thus, if $\mathcal{E} < 0$ the number of subsystems increases (subsystem size decreases) until $\mathcal{E} = 0$ where the free energy of the system is minimized (Eq. 10-116 in [22]), yielding the equilibrium nanocanonical ensemble. This $\mathcal{E} = 0$ is NOT the same as having negligible $\mathcal{E}$, characteristic of large homogeneous systems with no subdivisions. How and why $\mathcal{E} = 0$ is needed for stability in the generalized ensemble is discussed at length in Section 10-3 of [22]. Indeed, from Pg. 101: "Although $\mathcal{E}$ is negligible for a macroscopic system (…), it is *not* equal to zero in the strict sense that we are using $\mathcal{E} = 0$ above as an equilibrium condition. The macroscopic state is therefore not to be confused with the equilibrium state." To paraphrase: the homogeneous macroscopic state is not to be confused with the heterogeneous equilibrium state. An analogy is when the chemical potential is set to zero ($\mu = 0$) in standard statistical mechanics to ensure that extraneous restraints do not alter the distribution of quantized waves, e.g. photons and phonons. Likewise, $\mathcal{E} = 0$ ensures that extraneous restraints do not alter the average size and stable distribution of subsystems. Much of Hill's work in the early 1960's addressed the general foundations for finite-size thermal effects, but already by 1964 he recognized that the stable equilibrium requires $\mathcal{E} = 0$. Despite the rigorous details of Hill's initial discussions, since 1964 the stability condition seems to have been addressed only in the context of nanothermodynamics of subsystems inside large systems, not even in Hill's subsequent work on small-system thermodynamics [29-31].

Finding the stable equilibrium of independent subsystems inside larger systems involves taking the limit $\mathcal{E} \to 0$. This limit can be contrasted to the usual thermodynamic limit of standard statistical mechanics. The standard thermodynamic limit involves extrapolating to infinitely many particles ($N \to \infty$) and infinite volume ($V \to \infty$) while keeping constant density, $\rho = N/V$. This limit removes all finite-size effects from homogeneous systems, yielding the simplest (lowest-order) expressions for their behavior, but finite-size effects are essential to stabilize the thermal equilibrium and fluctuations inside most materials. Stable internal equilibrium also involves keeping a constant average density of particles in subsystems, $\bar{\rho} = \bar{n}/\bar{v}$, from their average volume ($\bar{v}$) and number of particles ($\bar{n}$). However, unlike $N \to \infty$, extrapolating $\mathcal{E} \to 0$ often yields small subsystems ($\bar{n} \sim 1$), so it might be called the "nanothermodynamic limit". In simple models where $\mathcal{E}$ can be calculated analytically, $\mathcal{E} = 0$ is used to fix the average size and distribution of subsystems. Alternatively, the stability condition of $\mathcal{E} = 0$ can be deduced by adjusting other parameters (e.g. $\mu/kT$) to give best agreement with measurements or simulations, then assuming that the system is in its stable equilibrium to yield the resulting parameters.

The concept of stable internal subsystems was first applied to measurements of glass-forming liquids in 1999 [32], then in 2000 to the critical behavior of ferromagnets where the term "nanothermodynamics" first appeared [27]. The term "nanocanonical ensemble" first appeared in 2006 for stable ensembles of internal subsystems [28]. These terms have subsequently been used in other contexts [33,34]. Often, they are used synonymously with Hill's general theory of small-system thermodynamics and his generalized ensemble of small systems [35-42] (though the term "nanocanonical" was not coined by Hill and did not appear until after he retired from science [29]). At first it seemed superfluous to distinguish nanothermodynamics from Hill's original theory – despite their distinct physical pictures they have similar mathematics – but in recent years confusion has been created when the terms nanothermodynamics and nanocanonical ensemble are used for unstable small systems having $\mathcal{E} \neq 0$ (e.g. [38-42]) with no reference to, or clarification of the conceptual implications and experimental evidence for $\mathcal{E} = 0$. The confusion reached the point where it was stated in [43] that "Hill's nanothermodynamics … has never been measured experimentally" and "… it remains elusive to relate the ensemble-dependent subdivision potentials to any experimental observables." One way to reduce the confusion would be to adopt new terms. For example, stable nanothermodynamics could be used for the nanocanonical ensemble of internal subsystems having $\mathcal{E} = 0$, then unstable, frozen, or out-of-equilibrium nanothermodynamics could be used for small systems having $\mathcal{E} \neq 0$. However, we recommend returning to the original terms of small-system thermodynamics for the study of small systems, with nanothermodynamics reserved for stable subsystems inside large systems. This respects the historical record of published priority, while restoring the legacy of Hill's seminal work from 1964 that already recognized the crucial conditions for stable equilibrium in the generalized ensemble of small systems.

Thermal and dynamic heterogeneity inside bulk samples are experimental observables that can be related to nanothermodynamics. Such heterogeneity has been measured by several experimental techniques for the primary response in most types of materials, including liquids and glasses [10,11,13,44-47], spin glasses [12], polymers [14], and crystals [26,27]. Because standard thermodynamics starts by assuming that simple systems are "*macroscopically homogeneous* (and) *isotropic*" [48], one conclusion might be that none of these materials is a simple system. However, even the semi-classical ideal gas (the prototypical system of standard thermodynamics) requires the nanocanonical ensemble from nanothermodynamics to maximize its entropy in stable equilibrium [22,25], as described in section *3.1*, below. For all such systems, that subdivide into internal subsystems, the theoretical "replica trick" from small-system thermodynamics that is purposely avoided in [43], becomes an unavoidable physical reality in nanothermodynamics, as found by many measurements on most types of materials.

Nanocanonical behavior of internal subsystems can be calculated analytically for simple systems, such as the semi-classical ideal gas and 1D Ising model [25]. The main mechanism driving subdivision is a net increase in entropy for the nanocanonical ensemble. Figure 2 is a cartoon sketch showing how a simple system containing two indistinguishable ideal gas particles (top) decreases its total entropy when subdivided into canonical subsystems (middle row shows the single pair of allowed subsystems) but increases its total entropy in the nanocanonical ensemble (bottom rows show some of the many possible subsystems). Here, each box represents a subsystem of indistinguishable particles, with internal lines representing interfaces between subsystems. Although dashed lines are moveable and permeable, they must make the subsystems statistically independent for entropy to be additive in the nanocanonical ensemble. Such subtle interfaces may come from breaks in the quantum exchange symmetry [49] due to decoherence [50] or interactions [51]. The resulting nanoscale heterogeneity yields a novel solution to Gibbs' paradox (section *3.1*, below), and allows atoms to be distinguishable by their locations when separated by macroscopic distances with many intervening atoms, e.g. when atoms are on opposite sides of a large room. This nanocanonical solution also yields sub-additive entropies as required for a fundamental theorem in quantum mechanics [20,52].

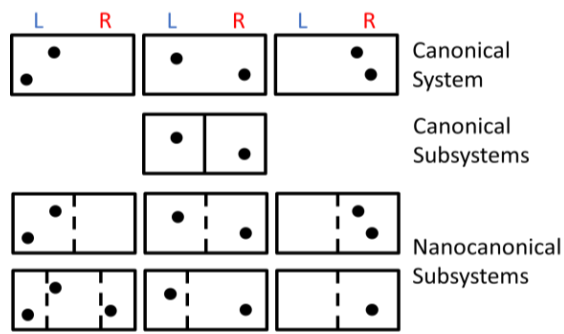

**Figure 2. Schematic representation of various multiplicities.** A canonical system (top) has two indistinguishable particles that may be on the left side, right side, or opposite sides. There is only one way to subdivide this system into canonical subsystems (middle), but there are many ways to subdivide it into nanocanonical subsystems (bottom). Adapted from Ref. [*25*]**.**

Most recent studies of small-system thermodynamics focus on evaluating computer simulations [36-42]. Usually, the simulations invoke a large reservoir to fix the thermodynamic variables, including $\varepsilon \neq 0$ characteristic of small systems that are not in stable equilibrium. These studies often utilize the "small system method" to deduce size-dependent effects by evaluating subvolumes (blocks) inside the simulation [53], with interactions between each block and its reservoir which yield the surface effects emphasized in these studies. However, a Hamiltonian of mean force is one way to show how a block that interacts directly with its bath "…is fundamentally different from the classical thermodynamic framework proposed by Hill…" [54], and therefore fundamentally different from nanothermodynamics for stable distributions of independent subsystems inside larger systems. In fact, a primitive form of the small system method having direct interactions across interfaces is the cellular method, introduced in 1949 to study local fluctuations in large systems of interacting particles [55]. Now it is known that such local fluctuations can deviate significantly from standard statistical mechanics, at least in MD simulations at low *T* (see [15] and Fig. 12, below), so that independent subsystems may be necessary for accurate modeling of real systems. Similarly, any correlations in the fluctuations of a block with its environment would fail the basic assumption of independent systems needed for Hill's theory [31] that is based on "a large sample of *independent* small systems (an 'ensemble')" (pg. 2 of [22], see also [48,56]). Indeed, a key requirement for any ensemble theory is that the "entropy is additive ($S_t = \eta S$) irrespective of whether the system is small or large" (pg. 13 of [22]). Thus, the key question for the relevance of nanothermodynamics is: does nature permit, and therefore prefer, internal heterogeneity from intermittent interactions? This heterogeneity is required for the stable nanocanonical ensemble of independent subsystems inside larger systems needed to minimize the free energy, as shown in Fig. 1. In any case, in the stable nanocanonical ensemble the reservoir yielding $\mathcal{E} = 0$ can come from neighboring subsystems that form a heterogeneous heat bath that is self-contained and self-consistent, facilitating the treatment of localized thermal behavior that is measured in most types of materials [10-14,25,26].

## 3. Brief review of some experimental evidence for nanothermodynamics

Most studies of nanothermodynamics focus on evaluating measured behavior, especially in systems that are in (or near) equilibrium [25-27,57,58]. In this case, the assumption of a homogeneous macroscopic system from standard thermodynamics ($\Delta F_t \approx 0$ in Fig. 1) is replaced by stable nanothermodynamics ($\mathcal{E} = 0$), yielding the nanocanonical ensemble of independent internal subsystems consistent with measured thermal and dynamic heterogeneity [10-14,25-27,44-46]. In fact, quantitative agreement is found between models based on nanothermodynamics and the measured temperature dependence of the magnetic correlation range ($\xi$) in cobalt [27], structural correlations in LaMnO$_3$ [26], and the typical sizes of independently relaxing regions in glass-forming liquids [44-46,57]. Furthermore, a mean-field cluster model that includes finite-size effects from the nanocanonical ensemble yields improved agreement with measured corrections to scaling near ferromagnetic phase transitions by maintaining $\mathcal{E} = 0$ as $\xi \to \infty$ [27]. Moreover, essentially the same model yields a gradual glass transition that is smeared out by finite-size effects when the correlation range is limited by disorder [32]. Finally, strict enforcement of the 2$^{nd}$ law of thermodynamics in nanoscale subsystems provides improved agreement with measured 1/$f$-like noise from qubits [25], metal films, spin glasses, and nanopores [58]. Here we briefly review six ways nanothermodynamics has been applied to simple models to yield improved agreement with measured behavior.

### 3.1. Novel solution to Gibbs' paradox

Gibbs' paradox refers to a prediction from classical statistical mechanics of ideal gas particles that violates the 2$^{nd}$ law of thermodynamics, first recognized by Gibbs in 1876 [59]. The violation occurs when an impenetrable barrier is moved into a box of distinguishable particles, subdividing the box into two halves, which reduces the volume that each particle can explore and reduces the total entropy. If the barrier is inserted reversibly (slowly and frictionlessly), this process lowers the entropy of a closed system, violating the 2$^{nd}$ law. The usual solution to Gibbs' paradox is to assume that all identical particles are statistically indistinguishable, as sketched in Figs. 3 A-C [25,59-64].

Figure 3 A shows a box that is initially subdivided into two sides, each having volume $V$ and $N$ particles. This initial state has type-X particles on the $L$ side (blue) and type-O particles on the $R$ side (red), with X's and O's indistinguishable particles (but they are distinguishable X's from O's). Each side of the box has the canonical ensemble (subscript $c$) partition function $Z_c = (V/\Lambda)^N/N!$. Here $\Lambda = h/\sqrt{2\pi mkT}$ is the thermal de Broglie wavelength, with $h$ Planck's constant and $m$ the mass of each particle, while the $N!$ removes overcounting of states assuming indistinguishable particles. In the canonical ensemble of small-system thermodynamics, the entropy of each side is given by the Sackur-Tetrode equation minus a non-extensive term that depends logarithmically (not linearly) on $N$ due to finite-size effects: $S_c(N,V) \approx Nk[5/2 + \ln(V/N\Lambda^3)] - k\ln(\sqrt{2\pi N})$. This non-extensive term comes from the subdivision potential, $\mathcal{E}_c \approx kT\ln(\sqrt{2\pi N})$, which cannot (and should not) be set to zero: mathematically because it comes from Stirling's formula for $N!$, and physically because it applies to the canonical ensemble of constrained (fixed size) subvolumes. This contribution to total entropy from small-system thermodynamics is negligible for large systems, $\mathcal{E}_c/NkT < 10$ ppm for $N > 10^6$, but non-negligible for small systems, $\mathcal{E}_c/NkT \approx 0.03$ for $N = 100$. Figure 3 B shows the box with the partition removed. The resulting increase in entropy is $\Delta S_{A \to B}(N,V) = 2S_c(N,2V) - 2S_c(N,V) = 2Nk[\ln(2)]$, as expected because each X and O has twice as much volume to explore. Figure 3 C shows the barrier returned to its original position, dividing the homogeneous mixture

of particles into two subvolumes. The change in entropy is $\Delta S_{B \to C}(N, V) = 4S_c(N/2, V) - 2S_c(N, 2V) = -k \ln(\pi N/2)$. Although this reduction in entropy depends logarithmically on the number of particles and is therefore relatively small for large systems, any violation of the 2$^{nd}$ law of thermodynamics could be a concern [60]. Furthermore, this entropy is super additive, failing a basic theorem of quantum mechanics [20,52]. Specifically, doubling the number of indistinguishable particles at constant density more than doubles the entropy: $S_c(2N, 2V) > 2S_c(N, V)$,

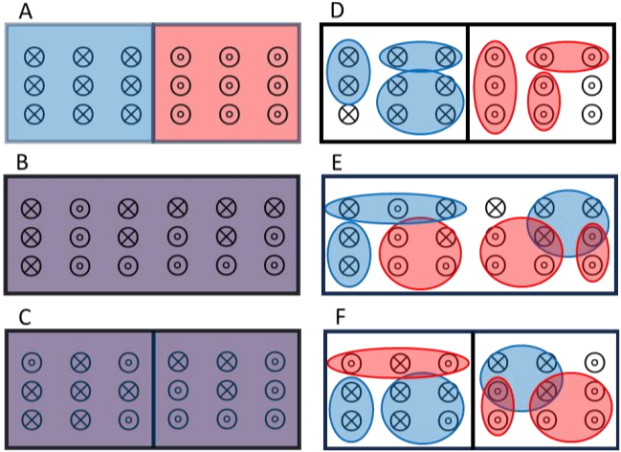

**Figure 3. Sketch showing two solutions to Gibbs' paradox for combining two types of particles: X's (blue) and O's (red).** (A)-(C) Canonical ensemble, where all particles of the same type are indistinguishable over all distances. (D)-(F) Nanocanonical ensemble, comprised of nanoscale subsystems, where similar particles can be distinguished by their location when in different subsystems. Adapted from [25].

Figures 3 D-F are sketches depicting a novel solution to Gibbs' paradox based on nanothermodynamics [25]. Figure 3 D shows the initial box with the same separation of particles as in Fig. 3 A, but now identical particles are indistinguishable only if they are close enough for exchange symmetry, yielding distinguishable particles when they are in distinct subsystems. In the nanocanonical ensemble, each subsystem has fluctuations in the number of particles and volume, so that only their average values are well defined, $\bar{n}$ and $\bar{v}$. The entropy of each subsystem is $S_{nc}(\bar{n}, \bar{v}) = \bar{n}k[5/2 + \ln(\bar{v}/\bar{n}\Lambda^3)] + k \ln(1 + \bar{n})$, which includes a Sackur-Tetrode-like equation plus a non-extensive term from subtracting the subdivision potential, $\mathcal{E}_{nc} = -kT \ln(1 + \bar{n})$. Assuming each side of the box in Fig. 3 D contains $\eta$ subsystems, the initial entropy of each side can be written as:

$$\eta S_{nc}(\bar{n}, \bar{v}) = \eta \bar{n}k[5/2 + \ln(\bar{v}/\bar{n}\Lambda^3)] + \eta k \ln(1 + \bar{n}) \tag{1}$$

Equation (1) consists of the Sackur-Tetrode equation (in square brackets) plus a non-extensive (final) term. Maximizing this final term maximizes the total entropy, as favored by the 2$^{nd}$ law of thermodynamics, and stabilizes the nanocanonical ensemble for internal subsystems. This final term also makes the entropy sub-additive, as needed for quantum mechanics [20,52]. Specifically, doubling the number of indistinguishable particles in an average subsystem at constant density reduces the entropy per particle: $S_{nc}(\overline{2n}, \overline{2v}) < 2S_{nc}(\bar{n}, \bar{v})$.

Equation (1) reveals the reasonable result that maximum entropy requires all three Legendre transforms: $E \to T$, $N \to \mu$, and $V \to p$. However, for large systems, the total increase in entropy from the first two Legendre transforms is negligible, with the final term in Eq. (1) coming entirely from the final transform into the nanocanonical ensemble. In fact, it is this final transform alone that causes significant deviations from the usual assumption that all ensembles yield equivalent results. Another issue in standard statistical mechanics that is solved by nanothermodynamics is the validity of ensembles having no extensive environmental variables. Indeed, Fig. 1 shows that thermal equilibrium requires the stability condition, $\mathcal{E}_{nc} = 0$. From the final term in Eq. (1), stability for the semi-classical ideal gas has $\bar{n} \to 0$, opposite to $N \to \infty$ for the usual thermodynamic limit of standard thermodynamics. However, both limits are taken in ways that maintain a constant density of particles, $\bar{\rho} = \bar{n}/\bar{v} = N/V = \rho$. One way to ensure uniform density for small subsystems in Fig. 3 D is to let the number of subdivisions in the ensemble diverge, $\eta \to \infty$, while fixing the total number of each type of particle ($N = N_t = \eta \bar{n}$) and the total volume of each box ($V = V_t = \eta \bar{v}$). Alternatively, by defining an average distance between particles, e.g. $\bar{a} = (\bar{v}/\bar{n})^{1/3}$ for each side of Fig. 3 D, then the local particle density ($\bar{n}/\bar{v} = 1/\bar{a}^3$) may vary in space and time as a system evolves because nanothermodynamics remains valid on the nanoscale.

There are two interesting limits in the non-extensive contributions to entropy from the nanocanonical ensemble, final term in Eq. (1). If all similar atoms are indistinguishable, as in the standard solution to Gibbs' paradox, each side initially has a single homogeneous subsystem $\eta \to 0$, so that the usual thermodynamic limit ($\bar{n} \to N \to \infty$) yields $-\mathcal{E}_{nc}/NkT = \ln(1+N)/N \approx 0$. In this case, the final term in Eq. (1) is negligible, leaving only the Sackur-Tetrode equation. If instead the system is in the nanothermodynamic limit where most atoms are distinguishable, $-\mathcal{E}_{nc}/kT = \ln(1 + \bar{n}) = 0$ yields $\bar{n} \to 0$ and $\ln(1 + \bar{n})/\bar{n} \approx (\bar{n} - \bar{n}^2/2)/\bar{n} \to 1$. Now the final term in Eq. (1) adds an extra entropy of $\eta \bar{n} k = Nk$ to the total, about 5.4% above the Sackur-Tetrode value for argon at standard temperature and pressure [25]. Indeed, the nanocanonical ensemble yields a significant increase in entropy (Eq. 1) and distinct physical picture (Figs. 3 D-F) for the semi-classical ideal gas, without changing the Hamiltonian. Thus, a full understanding of even the simplest models may require non-Hamiltonian contributions to energy contained only in Hill's fundamental equation of small-system thermodynamics shown in Fig. 1.

To test Eq. (1) against Gibbs' paradox we assume that $\bar{n}$ is vanishingly small but fixed by the stability condition, so that the starting value $\bar{n} = N_t/\eta$ in Fig. 3 D remains constant with density changes coming only from changes in $\bar{v}$. Upon removing the barrier, Fig. 3 E, the change in entropy is $\Delta S_{D \to E} = 2\eta S_{nc}(\bar{n}, \overline{2v}) - 2\eta S_{nc}(\bar{n}, \bar{v}) = 2N_t k[\ln(2)]$, matching the Sackur-Tetrode value. Now, if the barrier is returned to its original position, Fig. 3 F with $\eta \to 0.5\eta$, the entropy is precisely preserved for this reversible process: $\Delta S_{E \to F} = 4(0.5\eta)S_{nc}(\bar{n}, \overline{2v}) - 2\eta S_{nc}(\bar{n}, \overline{2v}) = 0$. Thus, Gibbs' paradox is solved with no violations of the 2nd law on any level, and without requiring indistinguishable particles beyond the nanoscale. Although Boltzmann provided an early solution to Gibbs' paradox by subdividing systems into cells of fixed volume [59,61], the nanocanonical ensemble justifies such subdivisions via increased entropy from variable volumes.

Ironically, experimental evidence for Eq. (1) may come from the accuracy of the Sackur-Tetrode equation for measured entropies. In [65], quantitative entropies of four monatomic gases are determined by adding all measured changes ($\Delta S_{>1}$) to a residual entropy that is assumed to be $S_1 \approx 0$ at $T = 1K$. This $\Delta S_{>1}$ is deduced from specific-heat and latent-heat measurements that yield the total change in entropy from the solid phase at $1K$ to the gas phase above the temperature of vaporization. For krypton, the values given in [65] (converted to units of $Nk$) are $S_1 = 0.00024Nk$ and $\Delta S_{>1} = 17.40Nk$, within 0.3% of the Sackur-Tetrode prediction, $S = 17.44Nk$. Similar measurements on neon, argon, and mercury yield values of $\Delta S_{>1}$ that are also within 0.07-1.4% of the Sackur-Tetrode predictions. However, the assumption of $S_1 \approx 0$ in [65] neglects known contributions to the residual entropy, such as the entropy of mixing from distinct isotopes that occur in natural abundances ($P_n$) [66,67]. Indeed, the stable isotopes of krypton and mercury should add extra residual entropies of at least $Nk \sum_n P_n \ln(1/P_n) = 1.229Nk$ and $1.280Nk$, respectively. Appendix A provides an explanation for the missing entropy in measured values of $S_1$, due to nanothermodynamics where most atoms remain distinguishable in both the liquid and gas phases.

### 3.2. The Ising model in stable equilibrium

The Ising model is the simplest microscopic model for a thermodynamic phase transition. As such, it remains among the most widely used models in statistical physics [68], providing basic insight into the thermal and dynamic behavior of many systems, and a simple guide to more sophisticated models. A solution to the one-dimensional (1D) Ising model was published by Ernst Ising in 1925, and a solution to the 2D Ising model was published by Lars Onsager in 1944, but neither solution is in stable equilibrium if finite-size effects are included. Here, two methods are used to give the stable solution to Ising's original model for a finite chain of interacting spins. Mathematical details are given in Appendix B.

The Ising model is based on binary degrees of freedom ("spins") with nearest-neighbor interactions that are usually attributed to quantum exchange. Figure 4 shows a specific system of $N = 9$ interactions between $N + 1 = 10$ spins in a 1D chain. For simplicity, let there be no external field. This model is readily solved using the statistics of the three types of interactions: low energy (●), high energy (**X**), and no energy (**O**) ("breaks"). These breaks, which must intermittently replace interactions for the equilibrium nanocanonical ensemble inside large systems, serve as the interfaces between subsystems. Mechanisms that could cause such breaks in real materials include Anderson localization or many-body localization [69,70]. Using $J$ as the ferromagnetic exchange constant, the net energy of interaction is $U = -J(N - 2N_x - N_0)$, where $N_x$ and $N_0$ are the number of high-energy interactions and breaks, respectively.

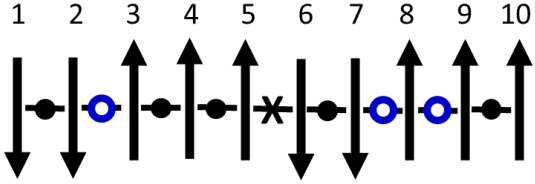

**Figure 4. Sketch showing a stable solution of the 1D Ising model at a given *T*.** Ten spins are in the chain. Each spin may be up or down. Each interaction between neighboring spins may be low energy (●), high energy (**X**), or a no-energy "break" (O) in the interaction.

Stable equilibrium in the nanocanonical ensemble comes from $\mathcal{E} = 0$, yielding $\Upsilon = 1$ in Eq. (B4) of Appendix B. An identical answer comes from Eq. (B10) for the average number of spins between breaks in the canonical ensemble of an infinite system, consistent with Fig. 4:

$$\bar{n} + 1 = 1 + \cosh(J/kT) \qquad (2)$$

There are multiple ways to confirm Eq. (2) for the thermal-average number of interactions in finite chains, $\bar{n} = \cosh(J/kT)$. First, $\bar{n} \to 1$ when $T \to \infty$, as expected for an equal mix of interactions (● or **X**) and breaks (**O**) that maximizes the entropy of mixing when energy can be neglected. Next, $\bar{n} \to \infty$ when $T \to 0$, as expected for an infinite homogeneous system that minimizes the energy when entropy can be neglected. Furthermore, as pictured in Fig. 4 and derived in Appendix B, Eq. (2) comes from adding breaks (**O**) to a large Ising system ($N \to \infty$), where nanothermodynamics subdivides the system into a stable distribution of subsystems, each having an average of $\bar{n} + 1$ spins. The same result comes from the generalized ensemble of small-system thermodynamics by treating finite chains of interacting (● or **X**) spins in a bath of spins having chemical potential ($\mu$) and temperature ($T$) in stable equilibrium ($\mathcal{E} = 0$), which yields an average of $\bar{n} + 1$ spins with breaks only at the endpoints of each chain. Thus, the nanocanonical ensemble of nanothermodynamics and the stable equilibrium of the generalized ensemble of small systems are equivalent, and both match the canonical ensemble of 1D spins with breaks.

Some additional conclusions come from the stable solutions of Ising's model given in Appendix B. A large system governed by a Hamiltonian with intermittent breaks (Eq. B6) solved in the canonical ensemble yields heterogeneous subsystems identical to Ising's original Hamiltonian (Eq. B1) in the nanocanonical ensemble, but only if the subsystems obey Hill's stability condition. Thus, distinct Hamiltonians can yield equivalent results when solved in different ensembles, again emphasizing the need to include all contributions to energy in Hill's fundamental equation of small-system thermodynamics shown in Fig. 1. Sections *3.3-3.6* utilize Ising-like spins in stable distributions of independent subsystems to improve the agreement between simple models and measured behavior.

*3.3. Mean-field cluster model for non-classical critical scaling*

Critical scaling is used to characterize divergent behavior near continuous phase transitions. One example is the divergence of magnetic susceptibility ($\chi$) in ferromagnetic materials near the Curie temperature, $T \gtrsim T_C$, which can be written as $\chi(T) \sim 1/(T - T_C)^\gamma$. Here we focus on the effective scaling exponent, $\gamma$, especially its temperature dependence. The "classical" (i.e. not modern) value from standard mean-field theory introduced by Weiss in 1907 has no temperature dependence, $\gamma = 1$, yielding the Curie-Weiss law that is measured at $T > 2T_C$. At $T < 2T_C$, however, $\gamma > 1$ is usually found. Although related non-classical critical exponents were routinely measured before 1900 [71], such measurements were mostly ignored until after 1944 when Onsager found a theoretical value of $\gamma = 7/4$ at $T_C$ in his solution of the 2D Ising model. Results from MC simulations of the 3D Ising model show a monotonic increase [72] from $\gamma \approx 1.1$ at $T \approx 3T_C$ to $\gamma \approx 1.24$ at $T_C$, but like Onsager's solution these simulations assume effectively infinite and homogeneous systems in the canonical ensemble, hence the simulations are not in the stable equilibrium of the nanocanonical ensemble. Consequently, many measurements fail to follow standard models, to the point where it has been said: "The critical exponents of iron and nickel are very similar to each other, while those for cobalt are clearly different. There is no theoretical understanding of these results." [73] And: "It is thus as if theorists and experimentalists in this field often behave like two trains passing in the night. Why is this?" [74]. Indeed, it might be said that ferromagnetic transitions should be added to the $\lambda$-transition in $^4$He [24,75] as examples where measured critical exponents remain unexplained by standard theories based on infinite and homogeneous systems. At least for ferromagnetic materials, a viable solution comes from adding heterogeneity that yields nanoscale subsystems, improving the agreement between measured critical behavior and standard theories, including real-space renormalization and Landau theory [76].

Here, we present a simplistic picture [26,27] for comparing three models of magnetic response, as sketched in Figs. 5 A-C. The magnetic susceptibility of these models at $T > T_C$ can be characterized by taking appropriate limits of:

$$\chi(T) = \frac{\chi_0}{\bar{n}\Gamma kT} \sum_{n=2}^{\infty} \sum_{\ell=0}^{n} \frac{2n!}{\ell!(n-\ell)!} (2\ell - n)^2 e^{\frac{n\left[\frac{J}{2}\left(\frac{2\ell}{n}-1\right)^2 + \mu\right]}{kT}} \tag{3}$$

The two sums in Eq. (3) are over the number of up spins ($\ell$) and the total number of spins ($n$) in the subsystems, starting at $n = 2$ to ensure at least one interaction per subsystem (although starting at $n = 0$ does not greatly alter the results). Note that $(2\ell - n)$ is proportional to the magnetic moment, which comes from the first derivative of the partition function ($\Gamma$) with respect to the magnetic field (that is set to zero and hence not included in Eq. (3)), while the susceptibility comes from the second derivative, yielding the factor of $(2\ell - n)^2$ in the summand. The adjustable parameters in Eq. (3) used to fit data are an amplitude pre-factor ($\chi_0$), energy scale ($J$), and the chemical potential ($\mu$).

The choice of model determines the behavior of Eq. (3). For the usual Ising model all spins are fully localized, with no spatial or temporal averaging, as sketched in Fig. 5 B. In this case, the mean-field energy $-\frac{J}{2}\left(\frac{2\ell}{n} - 1\right)^2$ and its multiplicity $\frac{2n!}{\ell!(n-\ell)!}$ are replaced by time-varying values having local correlations that preclude simple analysis in the 3D Ising model. At the other extreme, standard mean-field theory assumes averaging over a homogeneous and infinite system (i.e. $n \to \infty$ with no sum over $n$ in Eq. (3)), yielding the same average behavior for every spin, as sketched in Fig. 5 A. Fluctuations are infinitesimal for such large systems, yielding $\ell \approx n/2$ for $T > T_C$ so that the mean-field energy is negligible. Alternatively, the mean-field cluster model has mean-field behavior on nanoscale subsystems, as sketched in Fig. 5 C, which facilitates nanoscale fluctuations. This model is essentially a mixture of the two extremes: mean-field theory on small subsystems where time- and/or space-averaged behavior between nearby spins can yield simple values for the net behavior of each subsystem (see last two paragraphs of Section *3.5*, below).

Figure 5 D shows the temperature dependence of $\gamma$ from measurements on two magnetic materials (symbols, given in the legend) and the predictions of three models (lines) that are represented by the sketches in Figs. 5 A-C. The dotted (horizontal) line in Fig. 5 D at $\gamma = 1$ is from classical mean-field theory

based on assuming average behavior of all spins at every site, Fig. 5 A. The dashed line in Fig. 5 D comes from simulations of the usual 3D Ising model based on assuming a homogeneous system in the canonical ensemble with uniform exchange interaction between every neighboring spin, despite the assumption that each spin is localized to its own site, Fig. 5 B. Although a careful analysis of this model [72] confirms the monotonic increase in $\gamma$ with decreasing $T$ shown by the dashed line in Fig. 5 D, the data from ferromagnetic materials have a prominent peak in $\gamma$. A similar peak is also found in various critical fluids [77]. The limit $\gamma \approx 1$ as $T \to T_C$ is especially clear in detailed measurements on Gd [78].

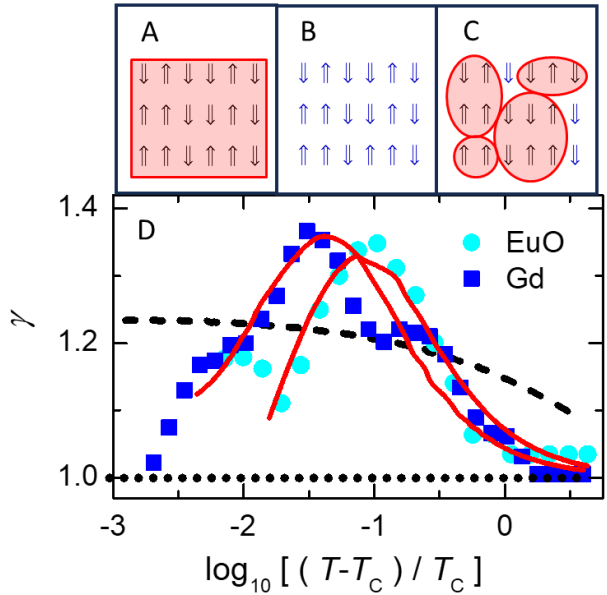

**Figure 5. (D) Temperature dependence of the effective scaling exponent from data (symbols) and models (lines) sketched in (A)-(C).** (A) Standard mean-field theory yields $\gamma = 1$ (dotted line in (D)). (B) Simulations of the standard Ising model yield a monotonic increase in $\gamma$ with decreasing $T$ (dashed line in (D)). (C) The mean-field cluster model yields non-monotonic behavior in $\gamma$ (solid lines in (D)), similar to measurements on EuO (circles) and Gd (triangles). Difficulty in determining $T_C$ yields uncertainty as $T \to T_C$, but not for $\log[(T - T_C)/T_c] > -2$ where $\gamma$ of the standard Ising model shows only gradual and monotonic behavior, unlike the measurements. Adapted from [26].

Solid lines in Fig. 5 D that mimic the measurements come from a mean-field cluster model [27]. Figure 5 C is a sketch indicating how this model is essentially a mixture of the models in Figs. 5 A and B. Specifically, mean-field energies are found for clusters of each size, then used as the effective energy of that cluster size. This cluster averaging allows local fluctuations (unlike averaging over the whole sample, Fig. 5A) and is consistent with the exchange interaction for spins that are delocalized or indistinguishable over the cluster (unlike spins localized to a single site, Fig. 5B). Equation (3) has two sums that transform the extensive environmental variables to two intensive variables, $\mu$ and $T$. The stability condition ($\mathcal{E} = 0$) connects these variables, yielding $\mu$ as a function of $T$. Often, $\mu/kT$ is relatively constant, e.g. for the ideal gas, $\frac{\mu}{kT} \sim -\frac{3}{2}\ln(T)$. The solid lines in Fig. 5 D come from fitting Eq. 3 using the constant value of $\mu/kT$ that gives best agreement with each set of data, as shown by the solid lines. This $\mu/kT$ from $\chi(T)$ then gives the $T$ dependence of other properties in the material, such as an average cluster size ($\bar{n}$) that mimics the measured temperature dependence of the magnetic correlation range in cobalt [27].

*3.4. A microscopic model for supercooled liquids and the glass transition*

Although amorphous materials comprise the oldest and most pervasive forms of synthetic substances, there is no widely accepted microscopic model for the mechanisms governing the glass transition. Thus, the glass transition is considered an "unsolved problem in physics" [24,79,80]; certainly, the standard Ising model is too simple to fully describe amorphous materials. However, by adding "orthogonal dynamics" to simulations of the Ising model on an equilibrium distribution of region sizes (similar to, but smaller than the clusters used for ferromagnetic materials in section *3.3*), a microscopic model is found that mimics at least 25 features measured in supercooled liquids and the glass transition [57]. It is called the orthogonal Ising model (OIM).

Orthogonal dynamics refers to time steps that separate changes in spin alignment ($m$) from changes in spin energy ($U$), so that the two relevant conservation laws can be uncorrelated if favored by the system. Energy-conserving changes in $m$ require spins that have no net interaction, e.g. spins 5, 6, or 8 in Fig. 4. Spin 8 ($+\rightarrow -$) because it has no interactions, spins 5 ($+\rightarrow -$) and 6 ($-\rightarrow +$) because they have an equal number of low- and high-energy interactions so that inverting the spin trades the energies without changing the net energy. Meanwhile, $m$-conserving changes in $U$ come from Kawasaki spin exchange (trading places of two interacting spins), which never changes the net $m$ but often changes the net $U$, e.g. spins 5 $\Leftrightarrow$ 6 in Fig. 4. Experimental evidence for orthogonal dynamics comes from observations that conservation of energy and conservation of (angular) momentum often occur on different time scales. One example is in magnetic resonance, where precession rates that change the spin alignment generally occur much faster than spin-lattice relaxation rates that change the energy. Other examples come from non-resonant spectral hole burning [10,12,14], where energy induced by a large-amplitude low-frequency pump oscillation can persist in local degrees of freedom for minutes, or even hours, while changes in dipole alignment often occur much faster. Insight into such separation of time scales may come from analyzing MD simulations [15], where equilibrium fluctuations often involve potential energy traded back-and-forth between neighboring spatial blocks for several atomic vibrations. Thus, conservation of local energy dominates over the relaxation of energy via slow coupling to distant parts of the sample that serve as the heat bath. Energy localization is found to require anharmonic interactions, presumably needed to scatter the harmonic modes (phonons) and cause the localized anharmonic modes to decouple from the phonons. In any case, orthogonal dynamics separates two main conservation laws allowing them to be independent if favored by the system, or to be correlated when appropriate. For a thermal transition that mimics liquid-glass behavior, the OIM is simulated utilizing the nanocanonical ensemble in 3D by adding Metropolis steps that make or break the interactions between neighboring spins [57].

Figure 6 shows frequency-dependent losses deduced from simulations of the OIM. This loss is found from the power spectral density ($PSD$, i.e. the magnitude squared of the Fourier transform of time-dependent fluctuations in $m$) using the fluctuation-dissipation theorem, with the normalization given in the label of the ordinate. The simulations are made on subsystems of two sizes, each at two temperatures, as given in the legends. Note that each loss spectrum from the larger subsystem ($n = 512$, green at $kT/J = 3.84$ or blue at $kT/J = 4.00$) shows three distinct maxima. The peak at lowest frequency is the primary response, the secondary response is at intermediate frequency, with the microscopic response at the highest frequency. The microscopic response involves inverting individual spins that have no net interaction with their neighbors, e.g. spins 5, 6, or 8 in Fig. 4, hence they invert with each attempt. The primary response comes from net inversions of the entire subsystem, yielding $m < 0$ to $m > 0$, or vice versa (see inset of Fig. 7 A), transiently connecting states that would have broken symmetry if not for finite-size effects. The secondary response comes from normal thermal fluctuations in $m$ and $U$ near one of the two free energy minima at $m \neq 0$. Thus, time-dependent changes in a single quantity $m$ in a microscopic Ising-like model yield all three main types of response found in supercooled liquids, but only if orthogonal dynamics is used to separate the three responses. As seen in Fig. 6, the simplicity of the model allows simulations over nearly ten orders of magnitude in frequency using a simple algorithm and relatively short computation times.

Response in the OIM involves changes in $m$ that are orthogonal to, but still influenced by fluctuations in $U$. Specifically, if the energy of a finite-sized subsystem fluctuates to $U$ from its average value $\bar{U}$, using $\delta U = U - \bar{U}$, a second-order Taylor-series expansion yields the change in entropy:

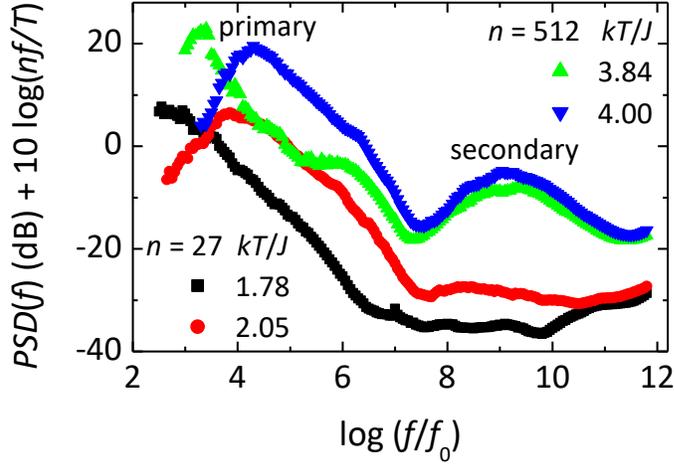

**Figure 6. Log-log plot of frequency-dependent loss from the orthogonal Ising model.** The loss is deduced from the power spectral density (PSD) using the fluctuation-dissipation theorem. The frequency is normalized by $f_0$ to put the microscopic peak at $\log(f/f_0) \sim 12$. Simulations are made on subsystems of two sizes, each at two temperatures, as given in the legends. Adapted from [57].

$$\delta S = \frac{\partial S}{\partial U}\delta U + \frac{1}{2}\frac{\partial^2 S}{\partial U^2}(\delta U)^2 = \frac{1}{T}\delta U - \frac{1}{2}\frac{1}{T^2 C_V}(\delta U)^2 \tag{4}$$

Here, definitions of temperature and heat capacity are used to give $\partial S/\partial U = 1/T$ and $C_V = \partial \bar{U}/\partial T$, respectively. The likelihood of such energy fluctuations is given by Boltzmann's probability, $p \sim e^{\delta S/k}$. Note that standard statistical mechanics is based on this probability applied to an infinite heat bath (subscript $H$), with negligible quadratic term and a negative linear term ($\delta S_H/k = -\delta U/kT$) from conservation of energy with the subsystem. Mean-field theory on the Ising model to second order in inverse subsystem size yields an analytic expression for the average energy [57]:

$$\bar{U} = -\frac{J}{2}\frac{T}{T - T_c}\left[1 - \frac{1}{n(1 - T_c/T)^2}\right] \tag{5}$$

Notice the similarity between the pre-factor in Eq. (5) and the Curie-Weiss law for magnetism with $T_c$ the Curie temperature, a consequence of mean-field theory on finite-sized subsystems that can fluctuate [27]. Furthermore [32], if this pre-factor is used as an activation energy in the Arrhenius law it yields the Vogel-Fulcher-Tammann (VFT) law, which is a common empirical formula for super-Arrhenius activation of the primary relaxation time ($\tau_\alpha$) in supercooled liquids. However, when a subsystem is in thermal equilibrium, this lowest-order term ($\frac{1}{T}\delta U$ in Eq. (4)) is balanced by the linear term of the heat bath ($\delta S_H$). Thus, dynamics in the orthogonal Ising model is governed by energy fluctuations, not activation. Using Eq. (5) in Eq. (4), with $\delta U \ll |\bar{U}|$ so that $U \approx \bar{U}$, Boltzmann's probability yields an expression for primary relaxation times that can be written as [57]:

$$\tau_\alpha = \tau_\infty \exp\left[\frac{1/C}{(1 - T_c/T)^2}\right] \tag{6}$$

Key parameters in Eq. (6) are the curvature coefficient $C \propto n$, and prefactor $\tau_\infty$ that gives the relaxation time of an infinitely large subsystem ($1/C \to 0$). The temperature-dependent divergence of the exponential argument in Eq. (6) is similar to the VTF law squared, hence the term "VFT2 law" is used for the behavior of Eq. (6).

Figure 7 A shows the $T_c/T$ dependence of $\log(\tau_\alpha)$ from models (lines) and measurements [81,82] (symbols) of glycerol, a glass-forming liquid. The inset of Fig. 7 A shows an interpretation of the response mechanism, described in the next paragraph. Figure 7 B shows the difference between the VFT2 function

(solid black line at origin), the measurements (symbols), and the VFT function (red line). Also shown is a fit to the MYEGA function (blue line), where slow dynamics is attributed to a double exponential instead of a finite transition temperature [83]. Figure 7 B is a type of Stickel plot [81] that utilizes a differential of $\ln(\tau_\alpha)$ as a function of $T_c/T$ (given in the ordinate label) that removes $\tau_\infty$ from Eq. (6), then takes the inverse 1/3 power to linearize the exponential argument with respect to $T_c/T$. The standard deviation between the measurements and VFT2 behavior of Eq. (6) shown in Fig. 7 B is at least an order of magnitude smaller than the other functions that each have one extra adjustable parameter. The inverse size dependence of the effective activation energy ($C \propto n$ in Eq. (6)) is similar to the size dependence known to give better agreement with the spectrum of response found in supercooled liquids [32,84]. Furthermore, values of $n$ deduced from fitting Eq. (6) to measurements and simulations of $\tau_\alpha$ [57] give good agreement with the sizes of dynamic heterogeneities measured directly by multi-dimensional nuclear magnetic resonance [44-46].

The inset of Fig. 7 A is a sketch of a free-energy double-well potential depicting the OIM interpretation of slow responses in amorphous materials. Note that due to the finite size of the subsystems, their free energies can have normal fluctuations represented by the arrows near the bottom of the left-side well, identified as the secondary response in Fig. 6. A key to understanding primary response comes from orthogonality: changes in $m$ never change $U$. Thus, inverting $m$ from one well to the other cannot come from simple activation over a fixed energy barrier, as evidenced by the rates near the alignment inversion [57]. Specifically, $U$ increases slowly up the barrier and rapidly down as expected for activated energies, but $m$ has opposite behavior: changing faster while $U$ increases then slower when $U$ decreases. Such non-activated slow response is attributed to energy fluctuations that open an entropy bottleneck, allowing the subsystem to find a pathway through the barrier, as represented by the zigzag path and arrow in the inset of Fig. 7 A. Thus, the OIM has complexity even in the relaxation of a single subsystem, from internal degrees of freedom that can alter the height and/or porosity of each barrier, not from activation of a single point in a fixed energy landscape. Additional complexity comes from a stable distribution of subsystem sizes due to the nanocanonical ensemble. In any case, orthogonal dynamics allows the equilibrium fluctuations of a single parameter ($m$) to exhibit all three types of response typically found in amorphous materials, Fig. 6.

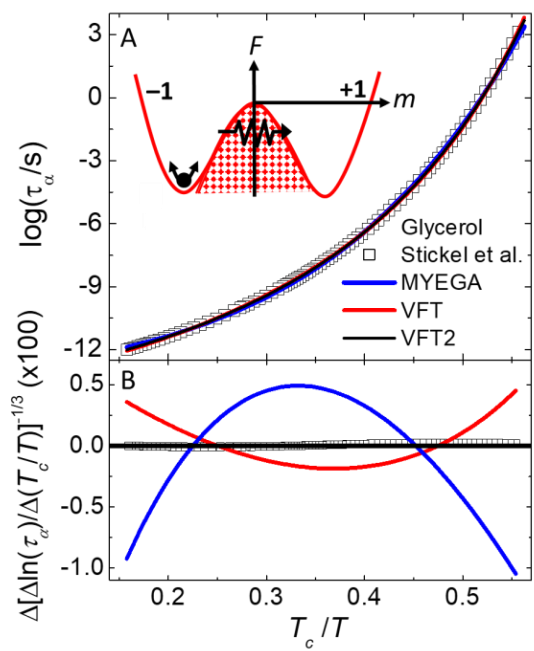

**Figure 7. Primary response time of glycerol.** Abscissa is inverse temperature, $T_c/T$, where $T_c$ is the mean-field critical temperature. The ordinate in (A) is $\log(\tau_\alpha)$, and in (B) it comes from a type of Stickel plot [81] utilizing finite differences of $\ln(\tau_\alpha)$, which remove the prefactor and linearize the VFT2 function. Symbols are from measurements (Stickel [*82*]). Various lines are from the VFT2 function Eq. (6) (black), VFT function (red), and MYEGA function (blue) [83]. The inset

is a sketch of a simple free-energy diagram, containing two minima separated by a barrier. Primary response in the orthogonal Ising model involves fluctuations in energy that open pathways between the minima. Adapted from [*57*].

*3.5. Maximum entropy as a mechanism for 1/f-like noise*

A general mechanism for 1/*f*-like noise comes from adding a local bath to internal subsystems, so that maximum entropy is maintained during reversible fluctuations by each subsystem plus its local bath. In the nanocanonical ensemble, the local bath may come from neighboring subsystems inside the larger system. Various models based on this mechanism exhibit not only 1/*f* noise, but also deviations from pure 1/*f* behavior that mimic measurements on thin metal films, nanopores, tunnel junctions and qubits [58,25]. Furthermore, when orthogonal dynamics is added, a 1D Ising model shows a crossover from 1/*f*-like noise to Johnson-Nyquist (white) noise at higher frequencies. Thus, although measurements (by Johnson) and theory (by Nyquist) of white noise were published together in 1927, it has taken nearly 100 years to find a model that simultaneously yields white noise and the 1/*f* noise that Johnson first reported in 1925.

Start by considering a system of two non-interacting and distinguishable binary degrees of freedom, e.g. two versions of spin 8 from Fig. 4 which can be either up or down. Let $m = (n_\uparrow - n_\downarrow)/n$ be the relative alignment, where $n_\uparrow$ and $n_\downarrow$ are the number of up and down spins, respectively. The multiplicities of the spin configurations, $W_m$, are $W_{+1} = 1$ for both spins up, $W_{-1} = 1$ for both spins down, and $W_0 = 2$ for the two ways that distinguishable spins can have one spin up and the other one down. Using the Boltzmann-Planck expression, $S_m = k \ln(W_m)$, the entropy of the two-spin system fluctuates up-and-down between $S_{\pm 1} = 0$ and $S_0 = k \ln(2)$. If this system was isolated from its environment, a fluctuation $k \ln(2) \to 0$ would violate the second law of thermodynamics. However, the system cannot be isolated because information about $m$ is needed to realize this entropy change. Otherwise, if the spin alignment is not (or cannot) be known, $m = ?$, the entropy of the system remains constant at $S_? = k \ln(4)$ from the $W_? = 4$ distinct configurations of the two spins. Assume that the spin system and its local bath form a closed system, and that the alignment of the spins is (or can be) sensed by the local bath, e.g. from the magnetic field on the bath. To avoid violating the second law of thermodynamics during reversible fluctuations, entropy from the system must be transferred back-and-forth to the local bath. Specifically, as the entropy of the system fluctuates up-and-down, the entropy of the local bath ($S_L$) must fluctuate down-and-up, so that total entropy remains maximized. Models for 1/*f*-like noise come from generalizing this idea of maintaining maximum entropy to interacting spins with $n \geq 2$.

Figures 8 A-E show all possible configurations of a subsystem containing $n = 4$ spins in a 1D chain. The figures are arranged in order of decreasing relative alignment, $m = +1$ in Fig. 8 A to $m = -1$ in Fig. 8 E. An exact expression for alignment entropy (preferrable to using Stirling's approximation, especially for small subsystems) comes from the binomial coefficient for the multiplicity of each alignment:

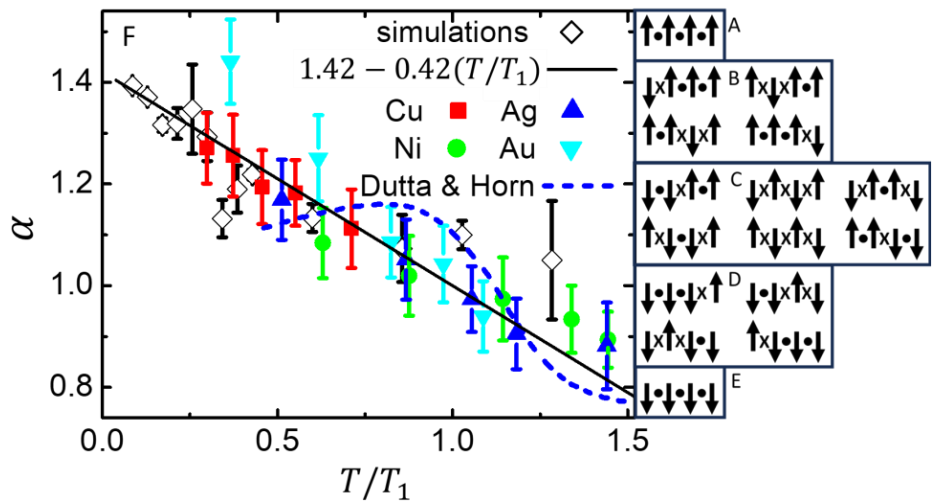

**Figure 8. 1/*f*-like noise from maintaining maximum entropy during equilibrium fluctuations.** (A)-(E) Sketch of all distinct configurations of $n = 4$ spins, arranged in order of decreasing alignment from $m = +1$ (top) to $-1$ (bottom). The multiplicity for the alignment entropy of the subsystem comes from the number of configurations in each box. (F) Temperature-dependent exponent for noise that varies as a function of frequency, $PSD(f) \propto 1/f^\alpha$, with the abscissa normalized by $\alpha = 1$ at $T_1$. Solid symbols (color) are from measurements [85] of noise in thin films for the metals given in the legend. Open symbols (black) are from simulations of a 3D Ising subsystems having $n = 27$ spins with dynamics usilizing a local bath to maintains maximum entropy during fluctuations in alignment. Solid line is the best linear fit to the simulations, weighted by the inverse variance of each point. Dashed line is from a random fluctuation model [86]. Adapted from [58].

$$S_m/k = \ln\left\{\frac{n!}{\left[\frac{n}{2}(1+m)\right]!\left[\frac{n}{2}(1-m)\right]!}\right\} \tag{7}$$

Note that if the subsystem fluctuates from a configuration with zero net alignment (Fig. 8 C) to all spins down (Fig. 8 E), its alignment entropy decreases from $S_0 = k\ln(6)$ to $S_{-1} = 0$. Thus, the entropy of the environment must increase an equal amount if violations of the 2nd law of thermodynamics are to be avoided. Assume that the environment consists of an ideal heat bath of entropy $S_H$ (that accommodates changes in energy) plus a local bath of entropy $S_L$ (that accommodates changes in alignment entropy of the subsystem). In nanothermodynamics these baths come from an ensemble of similar subsystems, so that together they form a self-consistent system with total entropy $S_T = S + S_H + S_L$. During equilibrium fluctuations the total entropy should be stationary with respect to all changes, which can be written as:

$$\Delta S_T = 0 = \left[\frac{\partial S}{\partial U}\Delta U + \Delta S_m\right] + \frac{\partial S_H}{\partial E_H}(\Delta E_H) + \Delta S_L \tag{8}$$

Here, the square brackets enclose terms for the entropy of subsystems, with $\Delta S_m$ from finite-size effects. Three conditions are implied by Eq. (8). The definition of temperature yields $\partial S/\partial U = \partial S_H/\partial E_H = 1/T$, conservation of energy requires $\Delta E_H = -\Delta U$, and maintaining maximum entropy during changes in alignment gives $\Delta S_L = -\Delta S_m$. Note that $\Delta S_m = S_m - S_0$ always involves the finite difference from $S_0$ because derivatives provide a poor approximation to the highly nonlinear changes in entropy during large fluctuations of small subsystems. Furthermore, the maximum entropy ($S_0$) remains the most-probable value for any subsystem that is above the critical temperature in zero external field.

Transition rates and probabilities for thermal fluctuations in energy and alignment are governed by Eq. (8) via the entropies of the baths. For example, the probability that the heat bath has an extra energy $\Delta E_H$ above its equilibrium value is $p_{\Delta E_H} \propto e^{\Delta S_H/k} = e^{\Delta E_H/kT}$. Thus, high-energy states are favored, opposite to the usual Boltzmann's factor, but necessary for the heat-bath entropy to yield Boltzmann's factor for subsystems using conservation of energy, $\Delta E_H = -\Delta U$. Indeed, it is this heat-bath entropy that is needed for Boltzmann's factor to give the statistics of systems in known states having no entropy. Similarly, the probability that the local bath has an extra amount of entropy $\Delta S_L$ above its equilibrium value is $p_{\Delta S_L} \propto e^{\Delta S_L/k}$, a term that is absent from standard statistical mechanics where contributions to the 2nd law from configurational entropy are usually neglected.

Transition rates come from applying detailed balance to each bath, $p_i r_{i\to j} = p_j r_{j\to i}$, where $r_{i\to j}$ gives the rate for changes from state $i$ to state $j$ in the bath. First, for transitions that increase the energy of the heat bath ($\Delta E_H > 0$): $\frac{r_{E_H \to E_H + \Delta E_H}}{r_{E_H + \Delta E_H \to E_H}} = \frac{p_{E_H + \Delta E_H}}{p_{E_H}} = e^{\Delta E_H/kT}$. The Metropolis algorithm for fast equilibration rates uses $r_{E_H \to E_H + \Delta E_H} = 1$, leaving $r_{E_H + \Delta E_H \to E_H} = e^{-\Delta E_H/kT}$. Thus, energy of the heat bath tends to increase faster than it decreases, opposite to the usual Metropolis algorithm, but needed for detailed balance of the multiplicities in the heat bath. Indeed, this is the fundamental mechanism for Boltzmann's factor and using $\Delta E_H = -\Delta U$ yields the Metropolis algorithm. Similarly, for transitions that increase the entropy of the local bath ($\Delta S_L > 0$): $\frac{r_{S_L \to S_L + \Delta S_L}}{r_{S_L + \Delta S_L \to S_L}} = \frac{p_{S_L + \Delta S_L}}{p_{S_L}} = e^{\Delta S_L/k}$. A Metropolis-like algorithm for fast equilibration rates uses $r_{S_L \to S_L + \Delta S_L} = 1$, leaving $r_{S_L + \Delta S_L \to S_L} = e^{-\Delta S_L/k}$. Using $\Delta S_L = -\Delta S_m = S_0 - S_m > 0$ to convert to the alignment entropy of the subsystem gives $r_{0 \to m} = 1$, leaving $r_{m \to 0} = e^{(S_m - S_0)/k}$. To reiterate, because of the highly nonlinear nature of the entropy difference of small subsystems, this $\Delta S_m$ in the exponent is the total entropy difference from the maximum-entropy configuration, not a differential.

By combining the usual Metropolis algorithm with the analogous $e^{(S_m - S_0)/k} > [0,1)$, where $[0,1)$ is a random number that is evenly distributed over the interval 0 to 1, simulations of the standard Ising model show $1/f$-like noise. In fact, using an exponent $\alpha \sim 1$ to characterize the power-spectral density, $PSD \propto 1/f^\alpha$, yields $1/f$-like noise with $\alpha$ that is often temperature dependent. Figure 8 F shows a comparison of $\alpha$ from measurements of noise from thin metal films [85] (solid symbols), a random fluctuation model [86] (dashed line), and MC simulations that maintain maximum entropy [58] (open symbols with error bars) with their weighted linear regression (solid line). The simulations are of the standard 3D Ising model on a simple-cubic lattice, with subsystems containing $n = 27$ spins. The measurements and simulations shown in Fig. 8 F are normalized by the temperature $T_1$ that gives $\alpha = 1$, with no other adjustable parameters.

Figures 9 A-E show sketches of a 1D chain of Ising spins, similar to those in Figs. 8 A-E. However, now there are $b = 4$ interactions (5 spins) arranged in order of decreasing energy per interaction: $u/J = (b_X - b_\bullet)/b$ from $u/J = +1$ (A) to $u/J = -1$ (E). Here $b_X$ is the number of high-energy interactions $(+J)$ and $b_\bullet$ is the number of low-energy interactions $(-J)$. The number of configurations in each figure represents the multiplicity of each energy (mirror-image configurations with the left-most spin up are not shown), yielding a multiplicity that is twice the argument of the logarithm in Eq. (7) with $m$ replaced by $u/J$. Thus in Eq. (8), if changes in alignment entropy ($\Delta S_m$) are replaced by changes in energy multiplicity ($\Delta S_{u/J}$), maintaining maximum entropy gives $1/f$-like noise in the energy fluctuations. Then, using orthogonal dynamics to allow decorrelated conservation laws yields fluctuations in alignment with an amplitude that is modulated by the energy. Specifically, Figs. 9 A ($u/J = 1$) and 9 B ($u/J = 1/2$) have a narrow range of alignments, $m = \pm 1/5$ (recall, configurations with the left-most spin up are not shown); Figs. 9 C ($u/J = 0$) and 9 D ($u/J = -1/2$) have an intermediate range of alignments, $m = \pm 1/5, \pm 3/5$; whereas Fig. 9 E ($u/J = -1$) has the broadest range of alignments, $m = \pm 1$.

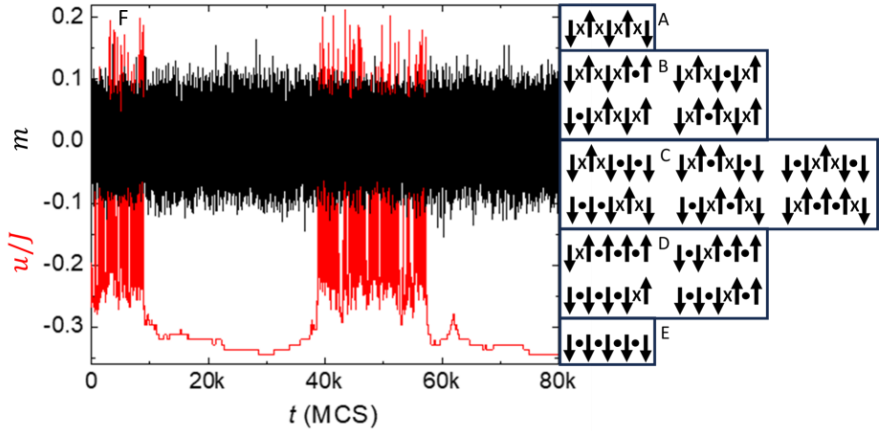

**Figure 9. Influence of energy on the amplitude of alignment fluctuations via orthogonal dynamics.** (A)-(E) Configurations of $b = 4$ interactions, arranged in order of decreasing energy. (F) Simulation of energy ($u/J$, red) and magnetization ($m$, black) as a function of time for the 1D Ising model containing $b = 1000$ interactions, with a local bath to maintain maximum entropy. Note how the amplitude of fluctuations in $m$ tends to be slightly larger when $u/J < -0.3$. Adapted from [25].

Figure 9 F shows a time series from a simulation of this model with $b = 1000$ (one contribution to the PSD in Fig. 10) [25]. Note how the fluctuations in alignment (black) tend to occur at a consistently fast rate, indicative of white noise, while the fluctuations in energy (red) exhibit jumps and steps over a wide range of time scales, indicative of $1/f$-like noise, also known as random telegraph noise or burst noise because if its appearance. Careful inspection reveals how the amplitude of fluctuations in $m$ is modulated by the value

of $u/J$. Specifically, fluctuations in $m$ tend to increase when $u/J < -0.3$, similar to how the range of alignments in Figs. 9 A-E increases with decreasing energy. Thus, normal fluctuations in alignment are dominated by white noise, with $1/f$-like noise in their second spectrum, consistent with measurements of thermal noise in equilibrium (non-driven) samples [87,88]. All these complexities come from a simple system based on the Ising model, with physically reasonable assumptions about orthogonal dynamics plus strict adherence to the 2nd law of thermodynamics to maintain maximum entropy. Thus, the single parameter ($m$) can exhibit both white noise and $1/f$-like noise, with no need for any separate distributions.

Figure 10 shows power spectral densities from simulations [25] (lines) and measurements [89] (symbols). The simulations are from the 1D Ising model with orthogonal dynamics and a local bath that maintains maximum entropy. Each PSD comes from $m$ as a function of time (e.g. black line in Fig. 9 F) by taking the magnitude squared of the Fourier transform of the time series. The PSD from simulations of smaller subsystems, e.g. $b = 50$ interactions (blue), show $1/f$-like behavior with bumps characteristic of individual Lorentzians, similar to the PSD from measurements of the flux noise ($\phi$) in a qubit (filled circles). The PSD of larger subsystems, e.g. $b = 1000$ interactions (red), show white noise at higher frequencies (dotted line) that crosses over to $1/f$-like noise at lower frequencies with an exponent of $\alpha \approx 0.92$ (dashed line). Most measurements exhibiting $1/f$-like noise show a similar crossover, including noise from the effective tunnel-coupling ($\Delta$) of a qubit (open circles). Thus, this single model provides a general mechanism for several features found in the measured noise from many systems.

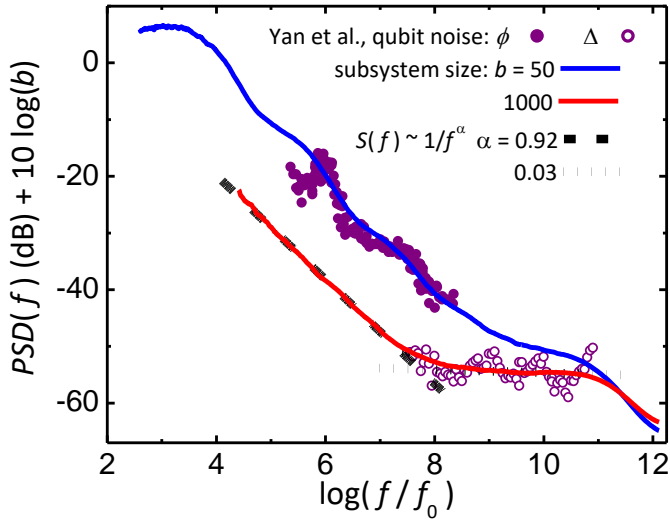

**Figure 10. Noise power spectral densities from simulations (lines) and measurements (symbols).** Solid lines are from fluctuations in alignment of 1D chains of $b + 1$ Ising spins using orthogonal dynamics while maintaining maximum entropy. Note that $b = 50$ (blue) is a small enough subsystem to show separate Lorentzians in a $1/f$-like spectrum, while $b = 1000$ (red) is large enough to show a crossover from white noise at high frequencies (dotted) to $1/f$-like noise at low frequencies with an exponent of $\alpha = 0.92$ (dashed). Symbols are from measurements of flux noise (solid) and tunnel-coupling noise (open) in a qubit [89]. Each set of measurements has been shifted in amplitude and frequency to match the simulations. Adapted from [25].

Another interesting aspect of orthogonal dynamics is that it can make mean-field theory exact for the 1D Ising model. In general, when averaged over all energies having a given alignment ($m$), the net energy of $n$ spins in a 1D chain ($n - 1$ interactions) is: $\bar{u}_m/J = 1 - nm^2$. Specifically, for the subsystems having $n = 4$ spins shown in Figs. 8 C-E: if $m = 0$ (Fig. 8 C) the average energy of the six configurations is $\bar{u}_m/J = (-1 + 3 + 1 + 1 + 3 - 1)/6 = 1$; if $m = -1/2$ (Fig. 8 D) the average energy of the four configurations is $\bar{u}_m/J = (-1 + 1 + 1 - 1)/4 = 0$; and if $m = -1$ (Fig. 8 E) the energy of the single configuration is $\bar{u}_m/J = -3$. All agree exactly with the mean-field expression $\bar{u}_m/J = 1 - 4m^2$. Indeed, if conservation of alignment in orthogonal dynamics persists long enough for the energy of a subsystem to exchange freely within its local environment before coupling to the heat bath, mean-field theory becomes exact in 1D due to averaging of all configurations having a given alignment. Thus, mean-field theory can be exact in 4+ dimensions due to spatial averaging, and in 1D due to time averaging. Although simple expressions have not been found for the Ising model in 2D and 3D, the accuracy of mean-field theory is generally improved when energies are time averaged using orthogonal dynamics. Such time averaging could also explain the good agreement between the mean-field cluster model and measured behavior of ferromagnetic materials shown in Fig. 5.

Time-averaging of 1D systems with orthogonal dynamics may also explain why mean-field theory mimics the measured properties of various linear-chain molecules, including biopolymers [90-92]. Indeed, we speculate that the behavior of linear-chain systems can be more-easily (and perhaps more-accurately) modelled by mean-field theory on short-segment subsystems, with a distribution of breaks in the exchange interaction between subsystems. Then, simple linear chains will have a stable equilibrium distribution of breaks, given by Eq. (2), while more-complex linear chains such as proteins can be modelled by mean-field theory on segment lengths that are shifted away from the equilibrium distribution by specific sequences of amino acids.

*3.6. The arrow of time in simple systems*

All basic laws of physics encountered in our daily lives are reversible except the second law of thermodynamics [93-97]. Indeed, it is this 2[nd] law that complicates the reconstruction of a raw egg that has fallen onto a hard surface, and prevents us from going back in time to stop it from falling. Sometimes it is said that there can be violations of the 2[nd] law [98-100], especially in small and simple systems, implying that it is a statistical rule of thumb valid only for large and complex systems. Further, it is often said that the explanation for 2[nd]-law behavior in complex systems comes from assuming that entropy increases for the vast majority of initial states, while entropy will decrease for some initial states. These assumptions have been tested and found to be false, at least for a simple (Creutz-like) model where the exact entropy can be calculated at every step for both the system and its heat bath [16]. Related behavior is found in molecular dynamics (MD) simulations of various models, which are based on Newtonian dynamics and hence intrinsically reversible [15]. Such deviations from standard statistical mechanics can be traced to the inherent reversibility and energy localization common to these simulations.

Simulations of the Creutz-like model and MD simulations corroborate a paradox in the theoretical properties of entropy [20], where systems with reversible dynamics cannot obey the 2[nd] law of thermodynamics. Specifically, entropy remains constant for classical systems due to Liouville's theorem, and for quantum systems due to Schrödinger's equation. According to [20], the only way to resolve this paradox and yield 2[nd]-law behavior in the dynamics is to utilize intrinsically irreversible (Markovian) steps, such as the master equation or Metropolis algorithm. Indeed, the simple Creutz-like model yields maximum entropy and 2[nd]-law behavior only if there is an intrinsically irreversible step [16].

The original goal of the Creutz model [101,102] was to utilize an explicit heat bath to avoid random numbers for efficient simulation of the 2D Ising model. For tests of the 2[nd] law the Creutz-like model starts with a 1D ring of *N* Ising-like spins, similar to the 1D chain in Fig. 4 but with periodic boundary conditions (which facilitates calculations of entropy) and with $2 \leq N \leq 2x10^6$. The heat bath consists of *N* Einstein oscillators, one oscillator per spin, but the spins often couple to distant oscillators to more-closely mimic an ideal heat bath. Each oscillator has an infinite number of equally spaced energy levels, $\varepsilon_i = iJ$, with $i$ a non-negative integer ($i \geq 0$) so that each oscillator acts as source of kinetic energy ("*ke* source"). The crucial and novel ingredient for the model to show the arrow of time is to include a thermal distribution of intermittent breaks, consistent with Eq. (2) and Fig. 4, and needed for stable equilibrium of the 1D Ising model.

The Creutz-like model is a type of cellular automaton [103,104]. Simulations of the model can be thought of as a type of microcanonical Monte-Carlo (MC) algorithm, where total energy is exactly conserved at every step. Dynamics in the model comes from choosing a site, then attempting to either flip the spin or change the interaction break-state for that site by exchanging energy with a *ke* source. If the change does not increase the potential energy (*pe*), it happens every attempt with any excess energy transferred to the *ke* source. If the change would increase the *pe*, it happens if and only if the *ke* source has sufficient energy, i.e. if the *ke* source can supply the potential energy and remain nonnegative. The entropy of the spins is given by the logarithm of their multiplicity, essentially the logarithm of the pre-factor to the exponent in the summand of Eq. (B7) (or Eq. (1) in [16]) comprised of the trinomial coefficient times $2^{N_0}$. The entropy of the *ke* bath is given by the logarithm of a binomial coefficient for the number of ways that the total kinetic energy from all *ke* sources can be shared among the *N* Einstein oscillators ([48] or Eq. (3) in [16]).

The Creutz-like model is simulated on systems of size $N = 2$ to $N = 2x10^6$ using various types of dynamics. Line color in Fig. 11 identifies dynamics that is reversible (black) or irreversible (red or green). All black and red lines in Fig. 11 come from simulations of large systems in the thermodynamic limit, $N = 2x10^6 \gg 10^4$, where $N \sim 10^4$ marks the crossover between behaviors, whereas the green lines in Fig. 11 F, from $N = 132 \ll 10^4$, show significant finite-size effects. In all cases the simulations have a fixed total energy ($E = NJ$) shared among all spins and *ke* sources. For reversible dynamics, each simulation sweep utilizes three randomly chosen (but fixed) sequences. One sequence gives the order of choosing each spin and its interaction. The second sequence gives the *ke* source for each spin-flip attempt. The third sequence gives the *ke* source for each break-state change attempt. Reversing the dynamics involves reversing every sequence. Such simulations are non-Markovian, having a long-term (but necessary) memory for the reversal. Irreversible dynamics is Markovian, utilizing a new random number for every choice of spin, interaction, and *ke* source.

Figures 11 A-E show the time dependence of the entropies per spin, $S/Nk$. Specifically: Fig. 11 C shows the entropies of the spin system, Fig. 11 B the entropies of the *ke* sources, and Fig. 11 A the total entropies (their sum). Even when the break-state attempt rate is reduced to 1/10 the spin-flip attempt rate (middle third of each simulation), with error bars visible if larger than the symbol size, Fig. 11 A shows that the total entropy from irreversible dynamics is several (> 4) standard deviations higher than the entropy of reversible dynamics. Figure 11 B mimics this difference, while Fig. 11 C mirrors the difference, showing that the increase in entropy is dominated by the *ke* bath. Thus, standard MC simulations of the Ising model will never show this entropy increase with intrinsic randomness, not only because the Metropolis algorithm is always Markovian, but also because the algorithm assumes essentially instantaneous coupling to an effectively infinite heat bath without explicit details about the coupling and how energy is conserved.

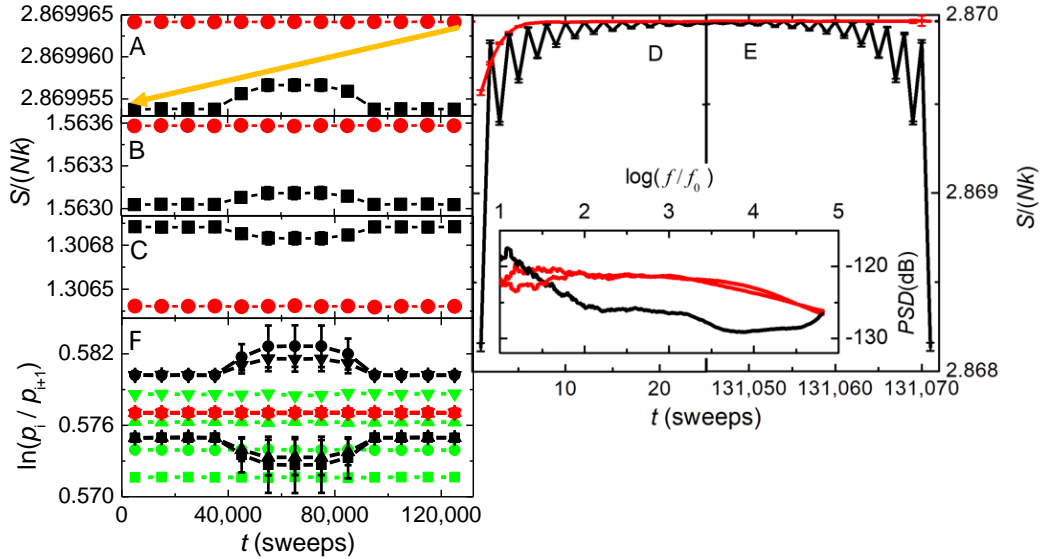

**Figure 11. Time dependence of entropies per *particle* (A)-(E) and inverse effective temperatures (F).** Simulations utilize a Creutz-like model of 1D Ising-like spins coupled to a *ke* bath of Einstein oscillators. Top three left-side graphs show the time-dependence of $S/(Nk)$ for the spins (C), *ke* bath (B), and their sum (A) in a large system, $N = 2x10^6$. Symbols come from first averag*ing* 10,000 sweeps, then averaging three separate simulations of each type, with error bars visible if larger than the symbol size. A simulation with irreversible dynamics (red circles) precedes every simulation with reversible dynamics (black squares). Thus, the total entropy always decreases when the dynamics becomes reversible, as indicated by the orange arrow in (A). Furthermore, when the rate of break-change attempts is reduced to 1/10 the rate of spin-change attempts (middle third of every simulation), reversible simulations have an entropy that depends on the dynamics. Right-side graphs show the total entropy, as in (A) but without time averaging, over a greatly expanded time scale. Here the differences between reversible (black) and irreversible (red) behavior are clearly visible at the start (D) and end (E) of the simulations. The inset shows corresponding differences in the power-spectral densities of the simulations. Symbols in (F) give the logarithm of the ratio of probabilities of neighboring energy levels in the *ke* bath, $\ln(p_i/p_{i+1})$, with $i = 0$ (squares), $i = 1$ (circles), $i = 2$ (up triangles), and $i = 3$ (down triangles). These values are proportional to the difference in inverse effective temperature of the adjacent levels. A single temperature applies only to irreversible dynamics in the thermodynamic limit (red), not for reversible dynamics in this limit (black) nor for irreversible dynamic of small subsystems, $N = 128$ (green). Adapted from [16].

Figure 11 F shows the time dependence of the logarithm of the ratio of occupation probabilities from neighboring energy levels in the *ke* bath, $\ln(p_i/p_{i+1})$. Symbol shape identifies the levels: $i = 0$ (squares), $i = 1$ (circles), $i = 2$ (up triangles), and $i = 3$ (down triangles). Therefore, if Boltzmann's factor applies to the statistics of the Einstein oscillators, Fig. 11 F gives the inverse of an effective temperature for neighboring energy levels, $J/kT_i$. Symbol color identifies the system size and type of dynamics. Green symbols show four distinct values of $J/kT_i$. Thus, even with irreversible dynamics, Boltzmann's factor fails to describe the *ke* bath of this system because it is not in the thermodynamic limit ($N = 128 \ll 10^4$), attributable to insufficient total energy in the microcanonical simulation to thermally occupy the higher levels. Black symbols show two (or more) distinct values of $J/kT_i$. Although this system is in the thermodynamic limit ($N = 2x10^6 \gg 10^4$), again Boltzmann's factor fails to describe the *ke* bath, but now because the dynamics is reversible. In contrast, overlapping red symbols show that irreversible dynamics in the thermodynamic limit ($N = 2x10^6$) yields a single temperature that is consistent with Boltzmann's factor.

The right-hand panels in Fig. 11 show the total entropy per particle at the start (D) and end (E) of the simulations. The data are the same as in Fig. 11 (A), but on an expanded time scale without time averaging. The red line in (D), from irreversible dynamics, shows that the initial entropy rises rapidly to a maximum value, then stays at this maximum to the very end of the simulation shown in (E). By contrast, the black line in (D), from reversible dynamics, oscillates with every sweep while evolving towards an entropy that is visibly less than the irreversible maximum entropy. Then, in (E) the system returns to its low-entropy initial state, exactly reversing every step in the process, even after nearly $10^{12}$ steps. This return to the initial state directly demonstrates Loschmidt's paradox for entropy that will decrease if reversible dynamics is inverted at a midpoint in time. The inset gives the power spectral densities of these simulations. Note how irreversible dynamics (red) shows a broad overdamped spectrum, with white noise below a characteristic frequency of $\log(f/f_0) \approx 3$. In contrast, reversible dynamics (black) has a sharp peak at the maximum frequency, corresponding to the oscillations seen in the main parts of these figures.

Insight into why systems having reversible dynamics deviate from standard statistical mechanics comes from details in the Creutz-like model. Recall that this model shows a difference between reversible and irreversible dynamics only if simulated with interactions that have intermittent breaks. Then, at least on short times with reversible dynamics, the breaks in the 1D system yield isolated segments of interacting spins that couple only to their own set of *ke* sources. Thus, energy tends to be localized within these subsystems, oscillating back-and-forth between the spins and their *ke* sources, causing the oscillations seen in Figs. 11 (D) and (E). In other words, due to intermittent interaction breaks, conservation of local energy overwhelms the slow transfer of energy to the rest of the system that serves as the large heat bath. However, when there is intrinsic randomness in the choice of *ke* sources, energy disperses quickly, without oscillations. Oscillations in energy also appear in molecular dynamics (MD) simulations of various models [15]. Like the reversible Creutz-like model with breaks, the oscillations can be attributed to inherently reversible Newtonian dynamics with anharmonic interactions that localize energy. Although these MD systems are too complex to yield entropy directly, their failure to follow standard statistical mechanics comes from excess fluctuations in energy that yield multiple local temperatures, similar to the black and green symbols in Fig. 11 F,

Figure 12 shows results from MD simulations of the Lennard-Jones (L-J) model in its crystalline phase at low $T$, adapted from Ref [15]. Symbol shape and color identify the simulation $T$, as given in the legend. (L-J units can be converted to physical quantities using values for specific substances, e.g. argon has $\varepsilon_0/k = 119.8$ K.) Simulations are made on a 3D system having 48 unit-cells on each side, yielding a total of $N = 4 \times 48^3 = 442{,}368$ atoms. Thus, the system forms a large heat bath for small internal blocks having $n \ll N$ atoms. (Blocks are subvolumes, without the interaction breaks needed for independently fluctuating subsystems. Indeed, neighboring blocks have correlated fluctuations as shown by the symbols in the insets.) The main part of Fig. 12 shows averaged and normalized values of the $pe$ fluctuations for blocks of $n = 32$ atoms as a function of the cutoff radius ($r_c$) for the L-J interaction. Using $pe$ as the potential energy per particle, standard statistical mechanics predicts $n(\Delta pe)^2/(kT)^2 = \partial(pe)/\partial(kT) = c_V/k$. Here, $c_V = 3k/2$ is expected for the specific heat from the equipartition theorem at low $T$ when the lattice is highly harmonic. Averaging the data from $kT/\varepsilon_0 \leq 0.1$ with $1.4 \leq r_c \leq 1.6$ yields $c_V/k = 0.92 \pm 0.03$. The overlapping symbols show the expected constancy with $T$ and $r_c$, but $c_V/k < 3/2$ due to harmonic modes that correlate the dynamics in neighboring blocks (see inset at $r_c = 1.5$). In contrast, $c_V/k$ becomes strongly $T$-dependent for $r_c > 1.6$, diverging as $c_V/k \propto 1/T$ when $T \to 0$.

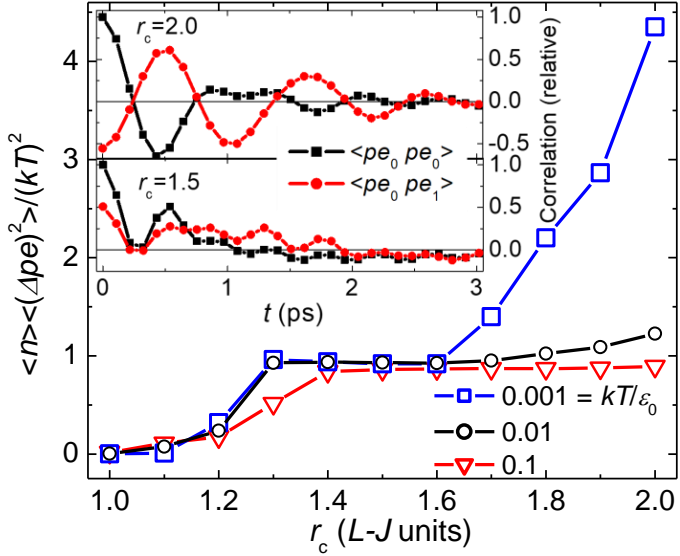

**Figure 12. Fluctuations in potential energy from MD simulations of Lennard-Jones crystals.** Main figure shows normalized $pe$ fluctuations for blocks of $n = 32$ atoms in a system of $N = 442{,}368$ atoms as a function interaction cutoff radius, $r_c$, at three temperatures given in the legend. Note that the data (open symbols) tend to be relatively constant (independent of $T$ and $r_c$) when interactions are robustly harmonic, having interaction between nearest-neighbor atoms only, $1.12 \approx 2^{1/6} \leq r_c \leq 2^{4/6} \approx 1.59$. Insets show the time dependence of energy autocorrelations in blocks (black squares) and energy correlations between nearest-neighbor blocks (red circles). Simulations are made at $kT/\varepsilon_0 = 0.0005$ for blocks containing a single unit cell of the crystal, $n = 4$. The lower inset shows that neighboring blocks are positively correlated when all atoms have robustly harmonic interactions ($r_c = 1.5$), while the upper inset ($r_c = 2.0$) shows that neighboring blocks are anticorrelated when interactions include 2nd-neighbor atoms that are anharmonic. Adapted from [15].

Quantitative agreement with $c_V/k \propto 1/T$ as $T \to 0$ is obtained by removing the 2nd-neighbor interaction from Boltzmann's factor [15]. These deviations from Boltzmann's factor can be attributed to conservation of local energy overwhelming the weak and slow coupling to distant blocks, due to energy that is localized by the intrinsic anharmonicity between 2nd-neighbor atoms that interact only when $r_c > 2^{4/6} \approx 1.59$. This interpretation is supported by the behavior of $pe$ correlations as a function of time shown in the two insets of Fig. 12. The insets show autocorrelations of $pe$ fluctuations within each block (black squares) and $pe$ correlations between each central block and its six nearest neighbors (red circles) for two different values of $r_c$. Note that in the purely harmonic crystal ($r_c = 1.5$, lower inset) *pe* fluctuations between nearest-neighbor blocks (red) are positively correlated. However, when there are second-neighbor interactions ($r_c = 2.0$, upper inset) neighboring blocks are strongly anticorrelated. In other words, when interactions extend beyond nearest-neighbor particles (upper inset), most of the *pe* in the fluctuation of each block (black squares) comes from a local bath comprised of nearby blocks (red circles), with negligible *pe* coming from the large heat bath of distant blocks. No wonder Boltzmann's factor fails to describe these fluctuations in *pe* that are localized to nearby blocks, effectively isolated from the heat bath. By contrast, in the purely harmonic lattice (lower inset), positive correlations are due to the dominance of plane-wave excitations that are positively correlated between neighboring blocks. These plane waves provide a uniform heat bath for the localized fluctuations, but only if the interactions are robustly harmonic. Similar deviations from Boltzmann's factor are found from MD simulations of other models having anharmonic interactions. In general, any system having realistic interactions and reversible dynamics will deviate from Boltzmann's factor, at least when observed closely enough.

Figures 11 and 12 show that computer simulations of various systems deviate significantly from standard statistical mechanics, but only if the dynamics is inherently reversible with energy fluctuations that are localized. Simple, intrinsically irreversible steps are sufficient for the Creutz-like model to yield maximum entropy and 2[nd]-law behavior. Common thermostats for MD simulations, such as the Nose-Hoover algorithm, are too weak to overcome conservation of local energy that causes deviations from Boltzmann's factor [15]. It is speculated [16] that in real systems, intrinsically irreversible dynamics may come from the quantum measurement process. Applying this idea to the Creutz-like model, when the spin system couples to its heat bath the specific amount of energy in the randomly chosen *ke* source is unknown until the coupling occurs, consistent with the calculation of entropy from the total energy that is shared among all *ke* sources. Future work is needed to explore this speculation, and whether intrinsically irreversible dynamics and a stable distribution of heterogeneous interactions can be added to MD simulations to improve agreement with standard statistical mechanics, and to measured behavior.

## 4. Summary

In 1964, Terrell Hill introduced the stability condition for his subdivision potential, $\mathcal{E} = 0$, a specific result in his general theory of small-system thermodynamics. Setting $\mathcal{E} = 0$ removes all extraneous restraints, yielding the stable equilibrium of small systems in the completely-open generalized ensemble. Since then, there has been little or no discussion of this stability condition in the literature for small systems, not even in Hill's subsequent work. Instead, starting in 1999 Hill's stability condition was adapted to become a central theme in the thermal equilibrium of subsystems inside larger systems. Indeed, $\mathcal{E} = 0$ is essential for independent internal fluctuations in real systems where it is difficult (if not impossible) to constrain the sizes of the subsystems. It was in this context that the term "nanothermodynamics" first appeared in 2000. Later, the related term "nanocanonical ensemble" was introduced for the stable distribution of subsystems inside larger systems. In this brief review we outline six basic problems that have been solved (or resolved) utilizing nanothermodynamics. We focus on how the stable nanocanonical ensemble allows simple models to give better agreement with measured behavior.

Section *3.1* and Appendix A show how using the nanocanonical ensemble provides a novel solution to Gibbs' paradox for the entropy of semi-classical ideal gases. The solution significantly increases the entropy above that of the Sackur-Tetrode equation, as favored by the 2[nd] law of thermodynamics. It makes the entropy sub-additive, as required by a theorem in quantum mechanics. It allows particles to be distinguishable by their location when separated by macroscopic distances, requiring particles to be indistinguishable only when close enough to have a quantum exchange. It confirms the accuracy of the Sackur-Tetrode equation for the measured entropies of several dilute gases. Finally, it precisely preserves the 2[nd] law of thermodynamics for systems of any size.

Section *3.2* and Appendix B describe the conditions for stable equilibrium of the original Ising model. In 1925, Ernst Ising solved the statistical mechanics of a 1D finite chain of $n + 1$ binary (uniaxial) spins with nearest-neighbor interactions. Although he ultimately let the number of spins diverge, $n + 1 \to \infty$, this thermodynamic limit gives thermal equilibrium only at $T = 0$. At nonzero $T$, and if the number of interacting spins is not artificially constrained, the stability condition ($\mathcal{E} = 0$) gives $\bar{n} + 1 = 1 + \cosh(J/kT)$ for the thermal-equilibrium average number of spins in a chain. Ising models in this stable equilibrium are used as a basis for the results in Sections *3.3-3.6*.

Section *3.3* describes a mean-field cluster model for non-classical critical scaling. This model can be pictured as a heterogeneous mixture of two homogeneous models for critical behavior, as shown in Fig. 5. Specifically, Curie-Weiss mean-field theory assumes that each spin exhibits the average behavior of all spins in the sample, yielding the classical critical scaling exponent of $\gamma = 1$ for $T > T_C$. Simulations of the 3D Ising model, solved in the canonical ensemble assuming an effectively infinite and homogeneous system with each spin localized to its site, predicts that $\gamma$ will increase monotonically with decreasing $T$ until $\gamma \approx 1.24$ as $T \to T_C$. However, measurements on various substances show that $\gamma$ increases to a maximum value of $\gamma \approx 1.35$ at about 10% above $T_C$, then decreases towards $\gamma \approx 1$ as $T \to T_C$. The mean-field cluster model, utilizing mean-field theory on a stable distribution of independent subsystems in the nanocanonical ensemble, mimics the measured behavior.

Section *3.4* describes the orthogonal Ising model (OIM) for the thermal and dynamic behavior of supercooled liquids and the glass transition. This model utilizes orthogonal dynamics, where changes in energy and changes in momentum (e.g. dipole alignment) never occur during the same time step, allowing these fundamental conservation laws to evolve on different time scales if favored by the system. The dynamics often exhibits multiple responses in a single parameter, consistent with the primary, secondary, and microscopic peaks measured in the dielectric loss of most supercooled liquids, as shown in Fig. 6. Applying mean-field theory to the OIM yields a novel expression (Eq. 6) for the $T$ dependence of the primary response time, $\tau_\alpha$. Because Eq. 6 is similar to the standard Vogel-Fulcher-Tammann (VFT) law, but with the square of reduced temperature in the denominator, it is called the VFT2 law. When compared to the VFT law, this VFT2 law gives at least an order of magnitude better agreement with measured values of $\tau_\alpha$ in various glass-forming liquids, even when analyzed in a way that the VFT2 law has fewer parameters, Fig. 7 (B). Such accurate agreement with the OIM suggests that slow dynamics in glass-forming liquids comes not from activation over energy barriers, but from equilibrium fluctuations in energy that facilitate passing through the barrier by opening pathways in an entropy bottleneck, as sketched in the inset of Fig. 7 (A).

Section *3.5* shows how various adaptations of the Ising model can mimic several features in the 1/*f*-like noise measured in the power-spectral density (PSD) of many systems. The key adaptation is to balance the entropy of the Ising system and the entropy of a local bath that maintains maximum entropy during equilibrium fluctuations. In general, maximum entropy is maintained by transferring any reduction in entropy of the system to an increase in entropy of the local bath. Thus, high-entropy states in the local bath favor low-entropy states in the system, similar to how Boltzmann's factor utilizes high-entropy states in the heat bath to favor low-energy states in the system. If there are no interactions between the spins, the resulting distribution of fluctuation times yield a PSD that varies as $1/f^\alpha$, with $\alpha = 1$. Adding interactions between spins allows the model to mimic other features from measured 1/*f*-like noise. One example is the $T$ dependence of $\alpha$ from voltage fluctuations in various metal films, Fig. 8. Another example, Fig. 10, utilizes a model with orthogonal dynamics that matches measurements of noise in a qubit. Specifically, the model gives Johnson-Nyquist (white) noise at high frequencies that crosses over to 1/*f*-like noise with $\alpha \approx 0.9$ at lower frequencies, and several Lorentzian-like steps from finite-size effects. Again, orthogonal dynamics allows a single parameter to show multiple features in the measured response. Furthermore, the stable nanocanonical ensemble of subsystems plus maximum-entropy mechanism for 1/*f*-like noise requires no adjustable parameters or ad hoc assumptions about distributions of environments to yield the behavior of fluctuation rates measured in many systems.

Section *3.6* provides new insight into the arrow of time using a simple model. The model consists of a 1D ring of Ising-like spins coupled to an explicit heat bath of Einstein oscillators, so that the total entropy of the system and its bath can be calculated at every step. By utilizing a randomly-chosen (but fixed) sequence of steps that couple the spins to their bath, the model can be made exactly reversible for unlimited numbers of steps. Alternatively, irreversible dynamics involves intrinsic randomness from a new random number for each step. Figure 11 shows that the total entropy is maximized only for intrinsically irreversible dynamics. Furthermore, if the system is initially in a maximum entropy state from irreversible dynamics, the total entropy always goes down when the dynamics becomes reversible. This is a clear counterexample to the common claim that the tendency for entropy to increase comes only from the rarity of initial states that cause entropy to decrease. Additionally, Fig. 11 F shows significant deviations from Boltzmann's factor for small systems, and for systems of any size with reversible dynamics. Thus, standard statistical mechanics applies only to large systems with intrinsically irreversible dynamics. The results can be attributed to how conservation of local energy often overwhelms the necessarily weak and relatively slow coupling to an effectively infinite heat bath. Finally, deviations from standard statistical mechanics in this simple system are consistent with the failure of Boltzmann's factor to describe energy fluctuations in MD simulations governed by inherently reversible Newton's laws, as shown in Fig. 12.

## 5. Conclusions

In this brief review we clarify how nanothermodynamics defines the stable equilibrium of small subsystems inside larger systems by minimizing free energy in the nanocanonical ensemble. Then we show how nanothermodynamics gives improved agreement between simple models and measured behavior. Examples include a mean-field cluster model for critical scaling in ferromagnetic materials, an orthogonal Ising model for the thermal and dynamic properties of liquids and glasses, and maximum entropy as a mechanism for 1/*f*-like noise in thin films and qubits. Nanothermodynamics also provides novel insight into some basic questions. One example is the thermal equilibrium of Ising's original model for finite chains of interacting spins in two different ensembles that start from two different Hamiltonians. Other examples are Gibbs' paradox for the entropy of ideal gases and Loschmidt's paradox for the arrow of time. It is anticipated that applying nanothermodynamics to other models may improve their agreement with measurements and provide new insight into nanoscale behavior on the inside of most materials.

**Acknowledgments:** We gratefully acknowledge contributions from Christian Arenz, Vincent B. Pizziconi, Steve Pressé, and George H. Wolf. We also thank Research Computing at Arizona State University for use of their facilities.

**Appendix A**

This Appendix suggests how having distinguishable particles in normal liquids and gases may explain the accurate agreement between measured entropies of dilute gases [65] and the nanocanonical expression for entropy, Eq. (1). (Another explanation is that unlike idealized semiclassical particles, real gas particles are indistinguishable beyond the nanoscale [25].) For Eq. (1) to agree with experiments, a constant residual entropy of $S_1 = Nk$ must be added to the measurements. This extra residual entropy would replace $S_1 \approx 0$ in [65] by altering the assumption that Debye's theory for the specific heat of normal solids extrapolates to $S_0 = 0$ as $T \to 0K$. No such assumption is necessary for measurements on $^4$He [105], where $S_0 = 0$ is assured because all atoms are indistinguishable when condensed into the superfluid ground state. Above $1K$ the measured entropy of $^4$He rises sharply until $\Delta S_{>1} \approx 0.76Nk$ at the start of the normal fluid phase, $T \approx 2.18K$. Nanothermodynamics provides an explanation for this value of $\Delta S_{>1}$ by treating most atoms as distinguishable in the normal fluid phase. Specifically, in the nanothermodynamic limit of the normal fluid phase where essentially all particles are distinguishable due to the nanocanonical ensemble of small systems, $\bar{N} \to 0$ with $N = \eta\bar{N}$. The non-extensive (final) term in Eq. (1) is $\eta k \ln(1 + \bar{N}) \to Nk$, yielding the residual entropy of $S_1 = Nk$. Now consider the superfluid phase of indistinguishable particles, in the thermodynamic limit of a large homogeneous system giving $\bar{N} \to N$ with $\eta \to 1$. In this case, the final term in Eq. (1) is negligible, $\ln(1 + \bar{N})/\bar{N} \approx 0$ so that $S_0 \approx 0$ as $T \to 0K$, consistent with the superfluid ground state. Because the normal fluid phase starts at $T \approx 2.18K$, the measured value of $\Delta S_{>1} \sim 0.76Nk$ implies that many (but not all) $^4$He atoms in the normal fluid phase are distinguishable. The standard solution to Gibbs' paradox, where all particles in the fluid and gas phases are indistinguishable, cannot explain this measured entropy of $^4$He, so that the excess entropy of the normal fluid would have to come from other sources. In addition, because the Gibbs' paradox solution from nanothermodynamics allows most atoms to be distinguishable in all phases, changes in entropy come primarily from changes in the volume explored by each atom as the system evolves from a solid to a gas.

In principle, most isotopically pure elements may form Bose-Einstein condensates in sufficiently dilute systems at low enough temperatures. Assuming that the entropy of such systems of indistinguishable atoms is $Nk$ lower than the entropy of a similar system of distinguishable atoms, this explains the value of $S_1 = Nk$ needed for Eq. (1) to give quantitative agreement with measurements of $\Delta S_{>1}$ in [65]. Furthermore, the nanocanonical ensemble provides an explanation for the measured excess entropy of $\Delta S_{>1} \sim 0.76 Nk$ found at $T \approx 2.18 K$ in the normal fluid phase of $^4$He.

**Appendix B**

In this Appendix, the stable equilibrium solution of the 1D Ising model is derived in two ways. Both ways yield finite chains of Ising-like spins. The Hamiltonians are of the form $U = -\sum_i J_i \sigma_i \sigma_{i+1}$, with $\sigma_i = \{+1, -1\}$ for {up, down) spins and $J_i$ the energy of interaction between neighboring spins, $\sigma_i$ and $\sigma_{i+1}$.

First, let all interaction energies have the same magnitude, $J_i = J$. The Hamiltonian for a finite chain of $n + 1$ spins ($n$ interactions), equivalent to Ising's original model in zero external field, is:

$$U_n = -J \sum_{i=1}^{n} \sigma_i \sigma_{i+1} \tag{B1}$$

Thus, there are two possible energy values for each interaction: $U_1 = +J$ or $-J$. The resulting potential energy of this system can be written as $U_n = -J(n - 2n_x)$, where $n_x$ is the number of $+J$ interactions between anti-aligned spins ($\sigma_i = -\sigma_{i+1}$). Using the binomial coefficient for the multiplicity of the two energy values of each interaction, the partition function in this finite-size canonical ensemble is:

$$Q_n = \sum_{n_x=0}^{n} \frac{2n!}{n_x!(n-n_x)!} e^{\frac{J(n-2n_x)}{kT}} = 2e^{\frac{nJ}{kT}}\left(1 + e^{-\frac{2J}{kT}}\right)^n$$
$$= 2[2\cosh(J/kT)]^n \tag{B2}$$

The average energy of this system is:

$$\bar{U}_n = \frac{\partial \ln(Q_n)}{\partial(-1/kT)} = -nJ\tanh(J/kT) \tag{B3}$$

Consider a finite spin chain that is in contact with a bath of spins, allowing the system to find its stable equilibrium distribution of sizes. The generalized-ensemble partition function ($\Upsilon$) comes from taking the final Legendre transform that removes the sole remaining extensive environmental variable ($n \to \mu$):

$$\Upsilon = \sum_{n=0}^{\infty} Q_n e^{\frac{\mu(n+1)}{kT}} = 2e^{\mu/kT}/[1 - 2e^{\mu/kT}\cosh(J/kT)]$$
$$= [e^{-\mu/kT}/2 - \cosh(J/kT)]^{-1} \tag{B4}$$

Although Ising's original paper includes a similar sum over all sizes, because his work preceded Hill's theory by about 40 years, he could not have known that the stable equilibrium for the nanocanonical ensemble is found by setting $\Upsilon = 1$. Except for [106], most textbooks neglect finite-size effects and treat "the Ising model" by assuming only a homogeneous system in the thermodynamic limit, $n \to \infty$. Instead, for the stable equilibrium of Ising's original model, set the subdivision potential to zero, $\mathcal{E} = -kT \ln(\Upsilon) = 0$ to give the chemical potential, $e^{-\mu/kT} = 2[1 + \cosh(J/kT)]$ and average number of spins:

$$\bar{n} + 1 = \frac{\partial \ln(\Upsilon)}{\partial(\mu/kT)} = [1 - 2e^{\mu/kT}\cosh(J/kT)]^{-1} = 1 + \cosh(J/kT) \tag{B5}$$

Note that the result of Eq. (B5) requires $n + 1$ as the factor for $\mu$ in Eq. (B4). This $n + 1$ is needed for the number of spins (not interactions) in each small system, and to ensure that every system in the sum from $n = 0$ has at least one spin, as needed for stability of the generalized ensemble [22]. Furthermore, although constant factors (such as 2 in Eq. (B2)) cancel from the average energy in Eq. (B3), every factor of 2 is needed to find the stable equilibrium in Eq. (B5), including the pre-factor of 2 from the degeneracy of a chain and its up-to-down mirror reflection. These results apply to the stable equilibrium of small systems in a uniform bath of heat and spins in the generalized ensemble of small-system thermodynamics, and to subsystems inside larger systems in the nanocanonical ensemble of nanothermodynamics, as shown in the next paragraphs.

Now consider the conceptually distinct but mathematically similar idea of a large system that subdivides into subsystems. Subdivision is accomplished by letting the interaction energy have two possible states: $J_i = \{0, J\}$ for {broken, made} interactions, yielding three possible values for each local interaction energy $U_1 = +J, -J,$ or $0$. The resulting Hamiltonian for a system of $N$ interactions is:

$$U_N = -\sum_{i=1}^{N} J_i \sigma_i \sigma_{i+1} \tag{B6}$$

Although superficially similar to an Ising model with quenched site disorder [107], here it is the interactions (not sites) that are disordered, and the breaks change intermittently (are not quenched) so that the system can find its stable equilibrium distribution of interactions, $J_i$. The potential energy of this system can be written as $U_N = -J(N - N_0 - 2N_x)$, where $N_0$ is the number of breaks ($J_i = 0$) and $N_x$ is the number of $+J$ interactions between anti-aligned spins ($\sigma_i = -\sigma_{i+1}$). Using the trinomial coefficient for the multiplicity of the three possible states of each interaction, the canonical-ensemble partition function of the system is:

$$\begin{aligned} Q_N &= \sum_{N_0=0}^{N} \sum_{N_x=0}^{N-N_0} \frac{2N!\, 2^{N_0}}{N_x!\, N_0!\, (N - N_0 - N_x)!} e^{\frac{J(N-N_0-2N_x)}{kT}} \\ &= \sum_{N_0=0}^{N} \frac{2N!\, 2^N}{N_0!\, (N-N_0)!} \cosh^{N-N_0}(J/kT) \\ &= 2^{N+1}[1 + \cosh(J/kT)]^N \end{aligned} \tag{B7}$$

The average subsystem size can be found from the average number of breaks ($\overline{N_0}$) using:

$$N - \overline{N_0} = \frac{\partial \ln(Q_N)}{\partial \ln[\cosh(J/kT)]} = N \cosh(J/kT)/[1 + \cosh(J/kT)] \tag{B8}$$

Then, the probability of a break becomes:

$$p = \overline{N_0}/N = [1 + \cosh(J/kT)]^{-1} \tag{B9}$$

Note agreement with expected limits: $p \to 0.5$ as $T \to \infty$ yielding the maximum entropy of mixing when interaction energies are negligible, and $p \to 0$ as $T \to 0$ yielding a homogeneous infinite system when entropy is negligible. For mathematical simplicity, assume that the system is in the thermodynamic limit ($N \to \infty$), allowing subsystems of any size, $0 \le n < \infty$. Using percolation theory, the average size of the subsystems is found from the number of interactions between breaks:

$$\begin{aligned} \bar{n} &= \sum_{n=0}^{\infty} n(1-p)^n p^2 \Big/ \sum_{n=0}^{\infty} (1-p)^n p^2 \\ &= \frac{\partial \ln[\sum_{n=0}^{\infty}(1-p)^n]}{\partial \ln(1-p)} = (1-p)/p = \cosh(J kT) \end{aligned} \tag{B10}$$

Thus, there is exact agreement in $\bar{n}$ from Eqs. (B10) and (B5). The average energy for this ensemble of subsystems is:

$$\overline{U}_N = \frac{\partial \ln(Q_N)}{\partial(-1/kT)} = -NJ \sinh(J/kT)/[1 + \cosh(J/kT)] \tag{B11}$$

Thus, at all $T > 0$, the energy in Eq. (B11) is higher than the energy in Eq. (B3). This difference is expected because breaks are now included in the system. Equivalence between the models is confirmed by comparing the average energy per interaction, $\overline{U}_n/n$, from Eq. (B3) and:

$$\overline{U_N}/(N - \overline{N_0}) = -J \sinh(J/kT)/\cosh(J/kT) \tag{B12}$$

In summary, a large system governed by the Hamiltonian of Eq. (B6) solved in the canonical ensemble yields subsystems that are equivalent to small systems governed by the Hamiltonian of Eq. (B1), but only if the small systems obey Hill's stability condition. One conclusion is that stable equilibrium in the generalized ensemble of small-system thermodynamics is equivalent to the nanocanonical ensemble in nanothermodynamics. Another conclusion is that distinct Hamiltonians can yield equivalent results when solved in different ensembles, emphasizing the importance of including all contributions to energy in Hill's fundamental equation of nanothermodynamics given in Fig. 1.